\documentclass[twocolumn,aps,pra]{revtex4-2}

\usepackage{graphicx}
\usepackage{color}
\usepackage{amsmath}

\usepackage{amssymb}
\usepackage{mathrsfs}
\usepackage{amsfonts}
\usepackage{bm,bbm}

\usepackage{mathtools}

\usepackage{esint}

\usepackage[T1]{fontenc}

\allowdisplaybreaks

\begin{document}

\title{Quantum thermochemical engines}

\author{Ugo Marzolino}
\email{ugo.marzolino@ts.infn.it}
\affiliation{University of Trieste, I-34151 Trieste, Italy}
\affiliation{National Institute of Nuclear Physics, Trieste Unit, I-34151 Trieste, Italy}
\affiliation{Scuola Normale Superiore, I-56126 Pisa, Italy}


\begin{abstract}
Conversion of chemical energy into mechanical work is the fundamental mechanism of several natural phenomena at the nanoscale, like molecular machines and Brownian motors. Quantum mechanical effects are relevant for optimising these processes and to implement them at the atomic scale. This paper focuses on engines that transform chemical work into mechanical work through energy and particle exchanges with thermal sources at different chemical potentials. Irreversibility is introduced by modelling the engine transformations with finite-time dynamics generated by a time-depending quantum master equation. Quantum degenerate gases provide maximum efficiency for reversible engines, whereas the classical limit implies small efficiency. For irreversible engines, both the output power and the efficiency at maximum power are much larger in the quantum regime than in the classical limit. The analysis of ideal homogeneous gases grasps the impact of quantum statistics on the above performances, which persists in the presence of interactions and more general trapping. The performance dependence on different types of Bose-Einstein Condensates (BECs) is also studied. BECs under considerations are standard BECs with a finite fraction of particles in the ground state, and generalised BECs where eigenstates with parallel momenta, or those with coplanar momenta are macroscopically occupied according to the confinement anisotropy. Quantum statistics is therefore a resource for enhanced performances of converting chemical into mechanical work.
\end{abstract}

\maketitle

\section{Introduction}

Energy conversion is at the basis of fundamental natural phenomena, and ubiquitous or breakthrough technologies 
\cite{Cheong2018,HasanuzzamanRahim2020}.
A paradigmatic model is represented by heat engines that convert heat into work with the limitations implied by the second law of thermodynamics \cite{Callen1985,Bejan2016}.
Similarly, chemical engines transform several kinds of chemical energy into other forms of energy.
For instance, mechanical work, e.g. volume expansion or directed difts and rotations, is generated
from chemical potential grandients \cite{Chen1997,Parmeggiani1999,Schmiedl2008,Esposito2009,Seifert2011,Seifert2012,Hooyberghs2013},
from the splitting of chemical bonds (molecular motors)
\cite{Julicher1997,Schliwa2003} as in ATP hydrolysis,
from thermal diffusion modelled by Langevin and Fokker-Planck equations (Brownian motors) \cite{Hanggi2009,Golubeva2012,Seifert2012},
and from surface energy
in interface phenomena like tears of wine, beating oil lenses, and self-oscillating pendant droplets \cite{Krechetnikov2017}.
Recent technologically oriented applications are syngas production \cite{Agrafiotis2015}, capacitive deionisation for desalinisation of blackish water \cite{Moreno2018,Moreno2019}, CO$_2$ capture by carbonation-decarbonation cycles \cite{Liu2020}, solar-driven CO$_2$ reduction \cite{Shuai2021}, chemical looping for hydrogen production \cite{Xiong2020,Li2021} and for energy and carbon storage \cite{Yan2020,Chein2021}, low-grade heat harvesting \cite{Dai2020,deVivanco2020,Chen2021}, pyrolytic reactions to increase the efficiency of turbine engines \cite{Wang2021}.

Heat engines have been investigated also in the quantum domain both at thermal equilibrium \cite{Saygin2001,Saygin2001-2,Humphrey2002,He2002,Chen2003,Kim2011,Zagoskin2012,Bengtsson2018,Huan2018} and out of equilibrium \cite{Kieu2004,Quan2007,Henrich2007,Anders2013,Goold2016,Vinjanampathy2016,Benenti2017,Bhattacharjee2021,Landi2021,Pekola2021,Ciccarello2022}.
Quantum effects become relevant already in systems at the nanoscale and in modern nanotechnologies that exhibit intermediate behaviours between classical and quantum regimes
\cite{Hanggi2009,Grafe2015,Steel2021}. Considerable advances have been done to observe the intermediate regime by tuning system parameters \cite{Esteban2012,Li2018,Ding2020,Tebbenjohanns2020,You2021}.
Chemical engines are indeed realised at these size scales, where
quantum effects
improve the performance of
existing synthetic molecular motors \cite{Yamaki2009,Goychuk2016,Oruganti2018,Majumdar2021}.
Nanomotors based on quantum dots have been conceived for charge pumping \cite{Brouwer1998,Switkes1999,Esposito2009}, and for converting electric work into mechanical work with high efficiency \cite{BustosMarn2013,Bruch2018,Ludovico2018}.
Quantum statistics is also relevant for the metabolic activity and related diseases \cite{Demetrius2003,Demetrius2006,Demetrius2010,Jinich2014,Labas2017} and in solar energy conversion \cite{Brabec2003,Markvart2016}.
Therefore, investigating the impact of fundamental quantum features on small-sized engines enables us to understand the deep quantum regime of these machines and the transition to the classical regime.

\subsection{Overview of the paper}

Taking inspiration from thermodynamic cycles modeling heat engines, this paper focuses on quantum engines that convert the chemical work of a working substance into mechanical work, by means of energy and particle exchanges with
thermochemical sources that control external driving (e.g., temperature, chemical potential, volume or particle number).
These machines, henceforth called thermochemical engines, are analogous to heat engines, where heat, temperature, and entropy are replaced respectively by chemical work, chemical potential, and particle number. Nevertheless, the second law of thermodynamics does not prevent to transform all the supplied chemical work into mechanical work without waste.
Irreversibility due to finite-time dynamics is introduced through quantum master equations that generalise Langevin and Fokker-Planck equations of Brownian motors.

The results presented in this paper show that quantum degenerate gases as working substances
provide maximum efficiency, i.e. without energy waste, while small efficiency and mechanical work output are obtained in the classical regime.
Quantum degenerate gases also imply large output power for finite-time irreversible cycles that perturb quasistatic processes.
Particular attention is devoted to the roles of standard and generalised Bose-Einstein condensates (BECs).
Standard BECs occur when a macroscopic particle number occupies the ground state \cite{PethickSmith,PitaevskiiStringari}, and have been experimentally realised with ultracold atoms \cite{Anderson1995,Bradley1995,Davis1995} or molecules \cite{Jochim2003,Xu2003},
photons \cite{Carusotto2013}, and with quasiparticles (polaritons \cite{Deng2010,Carusotto2013,Tang2021}, magnons \cite{Safonov2013}, phonons \cite{Misochko2004,Nardecchia2018}) even at room temperature.
Generalised BECs consist in large occupation of effective low-dimensional gases in the presence of highly anisotropic confinement volumes \cite{Krueger1967,Krueger1968,Rehr1970,vanDenBerg1982,vanDenBerg1983,vanDenBerg1986-1,
vanDenBerg1986-2,Ketterle1996,vanDruten1997,Beau2010,Mullin2012}.
Generalised BECs describe liquid Helium in thin films \cite{Osborne1949,Mills1964,Khorana1965,Goble1966,Goble1967}, magnetic flux of superconducting rings \cite{Sonin1969}, gravito-optical traps \cite{Wallis1996,Zobay2004}, and have stimulated experimental advances with ultracold atoms
\cite{Gorlitz2001,Greiner2001,Esteve2006,vanAmerongen2008}.

The above experimental realisations, together with recent progress in quantum simulators \cite{Georgescu2014,Monroe2020}, represent a plethora of platforms to implement thermochemical engines that are optimised in the deep quantum regime.
Of particular interest at the nanoscale are
BEC implementations with plasmon polaritons in a lattice of metal nanoparticles, that show ultrafast condensation at the sub-picosecond scale
\cite{Hakala2018,Vakevainen2020}, and with magnon BEC in ferromagnetic nanostructures \cite{Johnson2009,Chakravorty2017}.
Implementations with atomic gases exchanging particles and with highly anisotropic confinement allow for new atomtronic components \cite{Amico2021} based on high performance energy conversion.

The quantitative aspects of this paper are shown for ideal homogeneous gases, but similar behaviours remain valid for more general models. Indeed, they rely on physical conditions on the chemical potentials, and on mathematical properties of the average particle number that persist in the presence of different trapping potentials, density of states, and interactions, as discussed later on.

The rest of the paper is organised as follows. Section \ref{engines} describes the general scheme of thermochemical engines. Subsections \ref{Carnot} and \ref{Otto} are dedicated to equilibrium machines, called isothermal chemical Carnot cycle and isothermal chemical Otto cycle following the aforementioned analogy between thermochemical and heat engines. The effect of irreversibility as a perturbation of quasistatic processes on the efficiency and on the output power is discussed in section \ref{irr}. Conclusions are drawn in section \ref{concl}, and technical details are provided in appendices.

\section{Thermochemical engines} \label{engines}

The state of the working substance in thermochemical engines at thermodynamic equilibrium
is determined by the conjugated couples $(P,V)$, $(T,S)$, and $(\mu,\mathcal{N})$, where $P$ is the pressure, $V$ the volume, $T$ the absolute temperature, $S$ the entropy, $\mu$ the chemical potential, and $\mathcal{N}$ the average particle number.
At thermal equilibrium the substance is described by the grandcanonical statistical ensemble, with density matrix $\varrho=e^{-\beta(H-\mu N)}/Z$, Hamiltonian operator $H$, particle number operator $N=\sum_k a_k^\dag a_k$ (k labelling an orthogonal set of system modes), $\beta=\frac{1}{K_B T}$ the inverse temperature, and $Z=\textnormal{Tr}\,e^{-\beta(H-\mu N)}$ the partition function.

The grandcanonical ensemble provides a more transparent treatment of the effects of quantum statistics, and accounts for statistical fluctuating energy and particle number.
Allowing also for statistical fluctuations of the volume, the relevant ensemble is the so-called $\mu PT$ ensemble \cite{Hill,Campa2018,Marzolino2021}, studied for small systems and nanothermodynamics \cite{Chamberlin2000,Hill2001,Hill2002,Qian2012,Chamberlin2015,Latella2015,Bedeaux2018}, that predicts equations of state equivalent to the grandcanonical ones. Therefore the optimal performances of quantum thermochemical engines, proved in the following for the grandcanonical ensemble, can be straightforwardly generalised to the $\mu PT$ ensemble.

Thermodynamic transformations are parametrized by the free parameters of the grandcanonical ensemble, i.e., $V$, $\beta$, and $\mu$. The other thermal quantities are determined by the partition function:

\begin{align}
& \beta PV=\ln Z \ , \\
& \mathcal{N}=\textnormal{Tr}(\varrho N)=\frac{1}{\beta}\frac{\partial\ln Z}{\partial\mu} \ , \\
& U=\textnormal{Tr}(\varrho H)=-\frac{\partial\ln Z}{\partial\beta}+\mu\mathcal{N} \ , \\
& S=-k_B\textnormal{Tr}(\varrho\ln\varrho)=(U+PV-\mu\mathcal{N})/T \ .
\end{align}
Quasistatic transformations, where the system is always at thermal equilibrium, are described by a curve in the parameter space. Examples considered here are transformations fixing two of the aformentioned free parameters \footnote{Quasistatic thermodynamic transformations without exchange of particles require to fix only one variable because of the lack of the conjugated couple $(\mu,\mathcal{N})$.}.

The variation of the internal energy during a thermodynamic process is $\Delta U=Q-W^{\textnormal{M}}-W^{\textnormal{C}}$, where $Q=\int T\textnormal{d}S$ is the heat, $W^{\textnormal{M}}=\int P\textnormal{d}V$ is the mechanical work due to volume variations, and $W^{\textnormal{C}}=-\int\mu\,\textnormal{d}\mathcal{N}$ is the chemical work due to particle exchanges.
Fluxes of these energy contributions are generated when the working substance is put in contact with thermochemical sources.
Consider thermodynamic cycles consisting of several strokes,
where
$\Delta U_j$, $Q_j$, $W^{\textnormal{M}}_j$, $W^{\textnormal{C}}_j$ are the energy exchanges during the $j$-th stroke, and $P_j$, $V_j$, $\mu_j$, $\mathcal{N}_j$, and $\rho_j=\mathcal{N}_j/V_j$ are the thermal quantities at the beginning of the $j$-th stroke. The product of a cycle, also called \emph{load}, is the total mechanical work, that is positive if it is done by the working substance on the surrounding. The energy supplied to the working substance is the absorbed chemical work, i.e., the positive contribution to the internal energy due to particle exchanges $W^{\textnormal{C}}_{\textnormal{in}}=-\sum_j W^{\textnormal{C}}_j\Theta(-W^{\textnormal{C}}_j)$ where $\Theta(\cdot)$ is the Heaviside function. The energy released to the sources, $W^{\textnormal{C}}_{\textnormal{out}}=\sum_j W^{\textnormal{C}}_j\Theta(W^{\textnormal{C}}_j)$, contribute negatively to the internal energy. The energy $W^{\textnormal{C}}_{\textnormal{out}}$ is released to sources different from those that supply chemical work: the released chemical work can be delivered back to the substance only with a further energetic cost, so that $W^{\textnormal{C}}_{\textnormal{out}}$ is the waste generated for bringing the substance at the initial condition after a cycle. Consistently and in analogy with heat engines, the \emph{thermochemical efficiency} is the ration between the load and the supplied energy \cite{Krechetnikov2017}:

\begin{equation} \label{eff}
\eta=\frac{W^{\textnormal{M}}}{W^{\textnormal{C}}_{\textnormal{in}}}=1-\frac{W^{\textnormal{C}}_{\textnormal{out}}+\Delta U-Q}{W^{\textnormal{C}}_{\textnormal{in}}},
\end{equation}
where $\Delta U$ and $Q$ are the internal energy and the heat variations, respectively, during a cycle.

When reversible cycles are considered,
the figures of merit are the load per volume, $W^{\textnormal{M}}/\max V$ and the efficiency \eqref{eff}
with the condition
$\Delta U=0$.
These figures of merit will be used to compare performances of quantum and classical engines. Moreover, assuming that the temperature is kept constant along the cycle, infinitesimal heat variations are exact differentials and so $Q=T\int_{\textnormal{cycle}}\textnormal{d}S=0$. Therefore, the efficiency for reversible isothermal cycles becomes

\begin{equation} \label{eff.rev}
\eta_{\textnormal{rev}}=1-\frac{W^{\textnormal{C}}_{\textnormal{out}}}{W^{\textnormal{C}}_{\textnormal{in}}}\ ,
\end{equation}
and the load per volume is

\begin{equation} \label{load.rev}
\frac{W^{\textnormal{M}}}{\max V}=-\frac{W^{\textnormal{C}}_1+W^{\textnormal{C}}_3}{\max V}\ .
\end{equation}
The maximum efficiency $\eta_{\textnormal{rev}}\to1$ is then achieved if
the chemical potential is non-negative when the particles are injected in the working substance and is negative when the substance particles decrease.
These conditions imply $W^{\textnormal{C}}_{\textnormal{out}}=0$, recalling that $W^{\textnormal{C}}=-\int\mu\,\textnormal{d}\mathcal{N}$ for each transformation.

Ideal and interacting fermionic gases can exhibit both negative (at low density or high temperature) and positive (at high density or low temperature) chemical potentials \cite{PethickSmith,PitaevskiiStringari,Giorgini2008}, so that the thermochemical sources can fix their signs in order to obtain $\eta_{\textnormal{rev}}=1$. Ideal bosonic gases face a similar situation with positive chemical potentials replaced by vanishing chemical potentials in BEC phases, since the non-negativity of energy eigenstate occupancies implies the non-positivity of the chemical potential.
Bosonic gases with interactions also achieve maximum efficiency since their chemical potentials range from negative to positive values.
Indeed, repulsing particles
approaching the BEC transition show positive chemical potentials, proportional to the interaction strength and to the density, under several approximations, like the Bogoliubov, Hartree-Fock and Thomas-Fermi approximations \cite{PethickSmith,PitaevskiiStringari,OlivaresQuiroz2011}, effective mean-field \cite{Zagrebnov2001,Bhuiyan2021} and hard-core models \cite{Dai2007}, and with scattering length much larger than the interparticle distance (unitary gases) \cite{vanHeugten2013}. Quantum van der Waals interactions with a hard-core potential \cite{Vovchenko2015,Redlich2016} increase the chemical potential for repulsive interactions or for small attractive interactions (see appendix \ref{vanderWaals}). Therefore, the efficiency is maximised ($\eta_{\textnormal{rev}}=1$) for both fermions and bosons in the deep quantum regime.

Classical gases, especially with repulsive interactions, can also exhibit both negative and positive chemical potentials \cite{Sevilla2012}.
Nevertheless, the effects of quantum statistics can be neglected when (see appendix \ref{app.class})

\begin{equation} \label{class}
\mathcal{N}\ll V\left(\frac{2mK}{\hslash^2\mathcal{N}}\right)^{\frac{3}{2}},
\end{equation}
where $\hslash$ is the Planck constant, $m$ is the particle mass, $K$ is the average kinetic energy of the substance, and $K/\mathcal{N}$ is an intensive quantity, e.g., $K/\mathcal{N}=3k_BT/2$ for classical gases.
The condition \eqref{class} bounds the mechanical work of reversible isothermal engines:
$W^{\textnormal{M}}=-W^{\textnormal{C}}=\int_{\textnormal{cycle}}\mu\,\textnormal{d}\mathcal{N}\ll\mathcal{O}(\max V)$, where $\max V$ is the maximum volume attained during the cycle (see appendix \ref{app.class}). Since quantum gases are not constrained by Eq.~\eqref{class}, quantum engines provide a load per volume \eqref{load.rev} much larger than classical engines at comparable masses and energy densities. Another consequence of the constraint \eqref{class} is that the chemical potential of the classical van der Waals gas is always negative, thus preventing maximum efficiency, as detailed in appendix \ref{app.class}, contrary to what happens with the quantum van der Waals gas (see appendix \ref{vanderWaals}).

The rest of the paper is dedicated to concrete cycles where
the efficiency $\eta_{\textnormal{rev}}$ is
maximised with quantum degenerate gases, while
small efficiency and load are obtained in the classical regime.
Ideal homogeneous gases are discussed, and similar behaviours
extend to
interacting models that do not limit the efficiency range (as those discussed above),
and in the presence of general trapping potentials and density of states \cite{deGroot1950,Oliva1989,Yan1999,MartinezHerrera2019,Momeni2020} that do not alter the qualitative behaviours of homogeneous gases (e.g., the monotonicity and the (un)boundedness of the average particle number discussed in appendix \ref{app.ideal}).

\begin{figure*}[htbp]
\centering
\includegraphics[width=\textwidth]{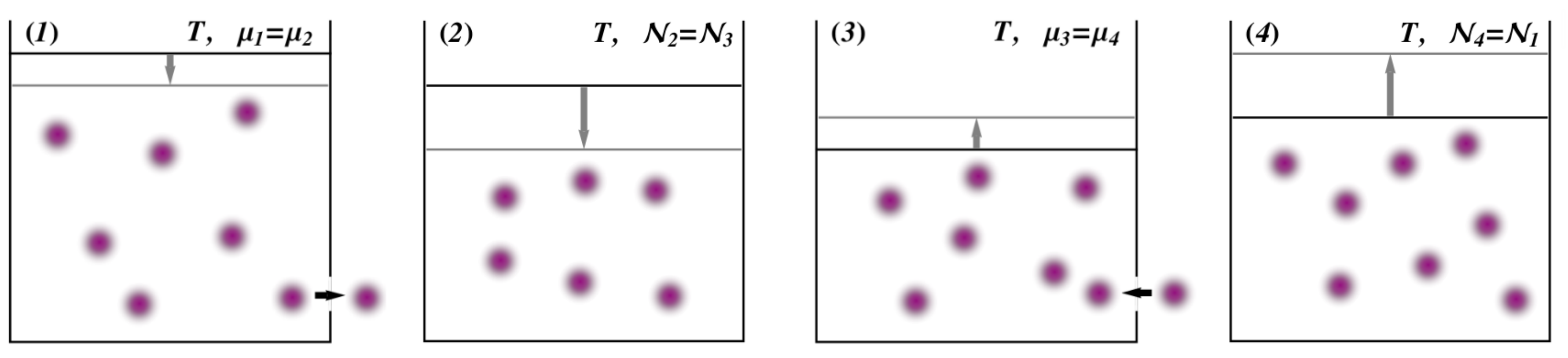}
\caption{
Pictorial representation of the isothermal chemical Carnot cycle.
}
\label{carnot.graph}
\end{figure*}

\subsection{Isothermal chemical Carnot cycle} \label{Carnot}

From the analogy with heat engines with the role of heat and temperature replaced by chemical work and chemical potential, the isothermal chemical Carnot cycle is defined by the following strokes:
\begin{itemize}
\item[\emph{(1)}] particle release at constant temperature $T$ and constant chemical potential $\mu_1=\mu_2$,
\item[\emph{(2)}] compression at constant temperature $T$ and constant particle number $\mathcal{N}_2=\mathcal{N}_3$,
\item[\emph{(3)}] particle injection at constant temperature $T$ and constant chemical potential $\mu_3=\mu_4$,
\item[\emph{(4)}] expansion at constant temperature $T$ and constant particle number $\mathcal{N}_4=\mathcal{N}_1$,
\end{itemize}
This cycle has the advantage to fix the chemical potentials when chemical work is done, as depicted in figure \ref{carnot.graph}, so that the condition for achievieng $\eta_{\textnormal{rev}}=1$ is directly controlled.

The performance of the isothermal chemical Carnot cycle depends only on the sign of the chemical potentials fixed by the sources, i.e., $\mu_{1,3}$, but not on the specific model, since $W^{\textnormal{C}}_1=-\mu_1(\mathcal{N}_2-\mathcal{N}_1)$ and $W^{\textnormal{C}}_3=-\mu_3(\mathcal{N}_4-\mathcal{N}_3)$. The supplied chemical work $W^{\textnormal{C}}_{\textnormal{in}}$, the released chemical work $W^{\textnormal{C}}_{\textnormal{out}}$, and the efficiency $\eta_{\textnormal{rev}}$ are respectively

\begin{align}
W^{\textnormal{C}}_{\textnormal{in}}= &
\begin{cases}
-W^{\textnormal{C}}_1=\mu_1\,(\mathcal{N}_2-\mathcal{N}_1) & \textnormal{if } \mu_{1,3}<0 \\
-W^{\textnormal{C}}_3=\mu_3\,(\mathcal{N}_4-\mathcal{N}_3) & \textnormal{if } \mu_{1,3}>0 \\
-W^{\textnormal{C}}_1-W^{\textnormal{C}}_3=W^{\textnormal{M}} & \textnormal{if } \mu_1<0 \textnormal{ and } \mu_3\geqslant 0
\end{cases}\ , \\
W^{\textnormal{C}}_{\textnormal{out}}= &
\begin{cases}
W^{\textnormal{C}}_3=\mu_3\,(\mathcal{N}_3-\mathcal{N}_4) & \textnormal{if } \mu_{1,3}<0 \\
W^{\textnormal{C}}_1=\mu_1\,(\mathcal{N}_1-\mathcal{N}_2) & \textnormal{if } \mu_{1,3}>0 \\
0 & \textnormal{if } \mu_1<0 \textnormal{ and } \mu_3\geqslant 0
\end{cases}\ , \\
\eta_{\textnormal{rev}}= &
\begin{cases}
\displaystyle 1-\frac{\mu_3}{\mu_1} & \textnormal{if } \mu_{1,3}<0 \\
\displaystyle 1-\frac{\mu_1}{\mu_3} & \textnormal{if } \mu_{1,3}>0 \\
1 & \textnormal{if } \mu_1<0 \textnormal{ and } \mu_3\geqslant 0
\end{cases}\ .
\end{align}
The load per volume,
\begin{equation} \label{load.carnot}
\frac{W^{\textnormal{M}}}{V_1}=(\mu_3-\mu_1)\left(1-\frac{\mathcal{N}_3}{\mathcal{N}_1}\right)\rho_1
\end{equation}
increases with the initial density $\rho_1$ and with the difference of chemical potentials fixed by the sources $\mu_3-\mu_1$.

Focus now on the isothermal chemical Carnot cycle working with ideal homogeneous gases
whose
thermal quantities are reported in appendix \ref{app.ideal}.
The classical limit holds for small fugacities $z=e^{\beta\mu}=\lambda_T^3\rho\ll1$, thus at negative chemical potentials and low densities. This condition implies small efficiency and load per volume.
On the other hand, fermionic gases do not have restrictions on the chemical potentials and densities, so that $\eta_{\textnormal{rev}}=1$ and arbitrary load per volume are achieved by fixing $\mu_1<0$ and $\mu_3\geqslant0$.
For bosonic particles, the chemical potentials are non-positive in order to have non-negative eigenstate occupancies, and approach zero at the formation of BECs. Therefore, maximum efficiency $\eta_{\textnormal{rev}}=1$, without restrictions on the load per volume, is attained if $\mu_3\to0$, namely when the working substance is a BEC during the third stroke.

\begin{figure*}[htbp]
\centering
\includegraphics[width=0.45\textwidth]{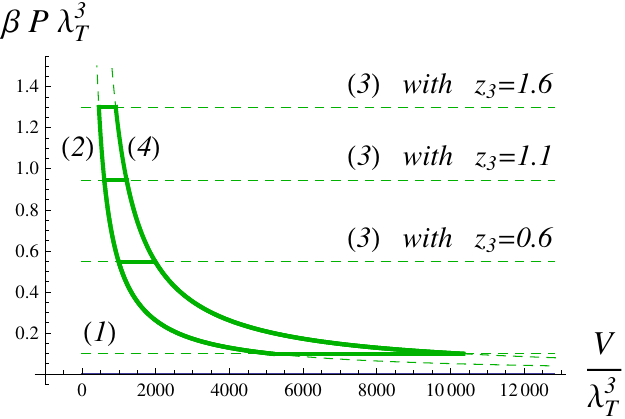}
\hspace{25pt}
\includegraphics[width=0.45\textwidth]{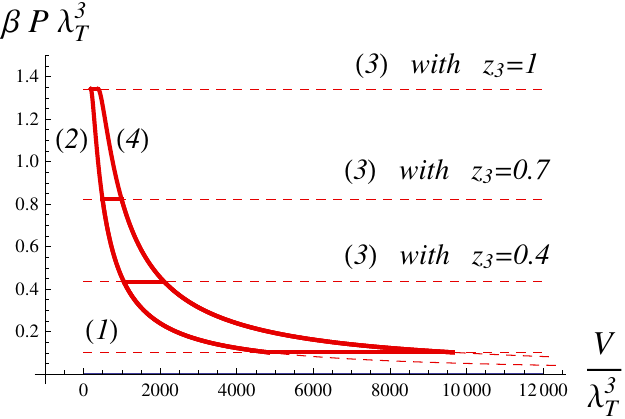}
\caption{
Rescaled pressure-volume diagram of the isothermal chemical Carnot cycle for fermions (left panel, green) with $\mathcal{N}_1=1000$, $\mathcal{N}_3=500$, $z_1=e^{\beta\mu_1}=0.1$ and $z_3=e^{\beta\mu_3}=0.6, 1.1, 1.6$, and for bosons (right panel, red) with $\mathcal{N}_1=1000$, $\mathcal{N}_3=500$, $z_1=e^{\beta\mu_1}=0.1$ and $z_3=e^{\beta\mu_3}=0.4, 0.7, 1$.
The numbers in parentheses indicate the strokes of the cycle corresponding to the closest curve.
}
\label{PV.carnot}
\end{figure*}

Figure \ref{PV.carnot} shows the pressure-volume diagram for fermions (left panel, green) and bosons (right panel, red), rescaled using factors kept constant during the cycle in order to plot dimensionless variables. The line integral, namely the area, of the closed line in the pressure-volume diagram is the extensive load and increases when the third stroke is pushed towards the deep quantum regime, i.e. increasing $\mu_3$, in accordance with equation \eqref{load.carnot}.

The analysis so far has revealed that the performances of the isothermal chemical Carnot cycle are optimised when the substance is in the deep quantum regime during the particle injection (third stroke), while the classical regime greatly underperforms.

\begin{figure*}[htbp]
\centering
\includegraphics[width=0.32\textwidth]{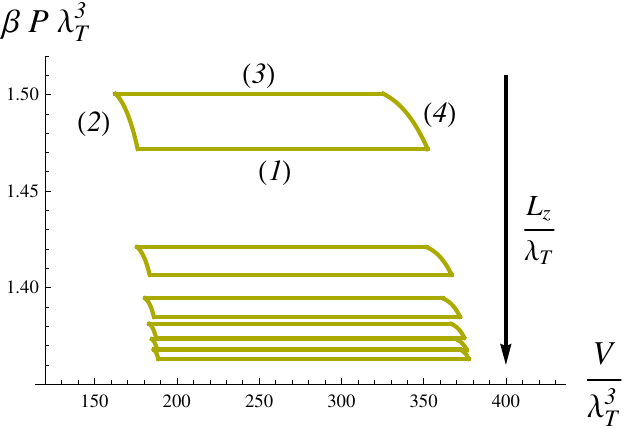}
\hspace{4pt}
\includegraphics[width=0.32\textwidth]{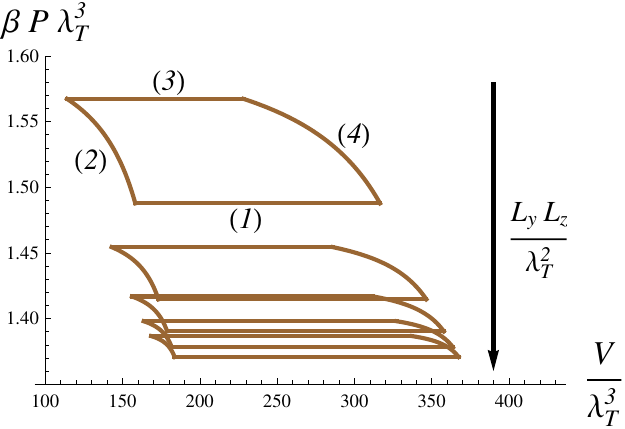}
\hspace{4pt}
\includegraphics[width=0.32\textwidth]{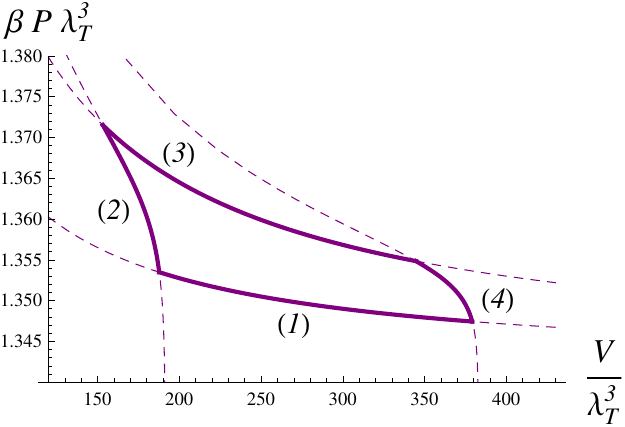}
\caption{
Rescaled pressure-volume diagram of the isothermal chemical Carnot cycle when the substance is always in a BEC with $\mathcal{N}_1=1000$, $\mathcal{N}_3=500$, $z_1=e^{\beta\mu_1}=0.9$, $z_3=e^{\beta\mu_3}=0.99$:
2D-BEC (left panel, yellow) with increasing $L_z/\lambda_T=10,20,30,40,50,60$ in the direction of the arrow,
1D-BEC (middle panel, brown) with increasing $L_yL_z/\lambda_T^2=10,20,30,40,50$ in the direction of the arrow,
and standard ground state BEC (right panel, purple).
The numbers in parentheses indicate the strokes of the cycle corresponding to the closest curve.
}
\label{PV.carnot.BEC}
\end{figure*}

When the substance is a BEC also during the first stroke, $\mu_1\to0$, and the efficiency depends on the geometry of the confinement volume $V=L_xL_yL_z$. This geometric effect on thermodynamic quantities of BECs is detailed in appendix \ref{app.ideal}. Consider now three different settings corresponding to standard ground state BECs, one- and two- dimensional generalised BECs, provided $L_x\geqslant L_y\geqslant L_z$.

If $L_x\sim L_y\sim L_z$ during the cycle, the BECs consist in the macroscopic occupation of the ground state (0D-BEC). The chemical potential scales as $\mu\simeq-(\beta f\mathcal{N})^{-1}$, where $f=1-\rho_c/\rho=1-(T/T_c)^{3/2}$ is the condensate fraction, $\rho_c$ is the critical density (i.e., the smallest density in the BEC phase), and $T_c$ is the condensation temperature (see Eq.~\eqref{Tc}).
The efficiency becomes, using also $\mu_1=\mu_2$ and $\mathcal{N}_2=\mathcal{N}_3$,

\begin{equation}
\eta_{\textnormal{rev}}=1-\frac{\mathcal{N}_2-\rho_cV_2}{\mathcal{N}_2-\rho_cV_3}\ .
\end{equation}

If $L_x\gtrsim\alpha' L_y L_z$, for a constant $\alpha'$, the BECs are effective one-dimensional gases consisting of states with momenta parallel to $L_x$ (1D-BEC). The chemical potentials are $-\beta\mu\simeq\pi(f\lambda_T\mathcal{N}/L_x)^{-2}$,
and the efficiency reads

\begin{equation}
\eta_{\textnormal{rev}}=1-\frac{(\mathcal{N}_2-\rho_cV_2)^2\,L_{x,3}^2}{(\mathcal{N}_2-\rho_cV_3)^2\,L_{x,2}^2}\ .
\end{equation}

If $L_y\gtrsim e^{\alpha L_z}\textnormal{poly}(L_z)$, where $\alpha$ is a constant and $\textnormal{poly}(L_z)$ stands for a polynomial in $L_z$, the BECs are effective two-dimensional gases made of states with momenta in the $x$-$y$ plane (2D-BEC). The chemical potentials are $-\beta\mu\simeq e^{-f\lambda_T^2\mathcal{N}/(L_xL_y)}$
and the efficiency

\begin{equation}
\eta_{\textnormal{rev}}=1-e^{(\rho_c-\rho_3)\,L_{z,3}\,\lambda_T^2-(\rho_c-\rho_2)\,L_{z,2}\,\lambda_T^2}
\end{equation}
approaches $1$ in the thermodynamic limit, e.g., if $L_{z,3}=L_{z,2}$ or $L_{x,3}L_{y,3}=L_{x,2}L_{y,2}$ recalling $V_3<V_2$.

If the substance is in a $d$D-BEC during the first stroke and in a $d'$D-BEC in the third stroke, with $d'<d$,
$L_x\gg\alpha' L_y L_z$ for the 1D-BEC, and $L_y\gg e^{\alpha L_z}\textnormal{poly}(L_z)$ for the 2D-BEC,
then $\eta_{\textnormal{rev}}\to1$ in the thermodynamic limit.

The rescaled pressure-volume diagrams are plotted in figure \ref{PV.carnot.BEC} for substances always in a 2D-BEC (left panel, yellow), in a 1D-BEC (middle panel, brown), and in a 0D-BEC (right panel, purple). The load, i.e., the area enclosed within the closed curve is subextensive but the efficiency can achieve large values as shown above.

\begin{figure*}[htbp]
\centering
\includegraphics[width=\textwidth]{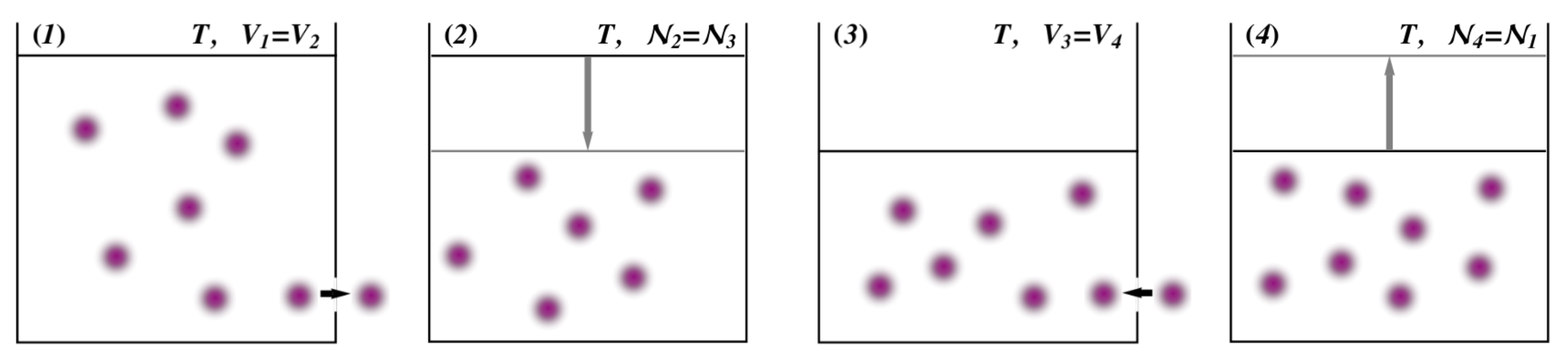}
\caption{
Pictorial representation of the isothermal chemical Otto cycle.
}
\label{otto.graph}
\end{figure*}

\subsection{Isothermal chemical Otto cycle} \label{Otto}

Following the aforementioned analogy between heat and thermochemical engines, the isothermal chemical Otto cycle consists in the following strokes:
\begin{itemize}
\item[\emph{(1)}] particle release at constant temperature $T$ and constant volume $V_1=V_2$,
\item[\emph{(2)}] compression at constant temperature $T$ and constant particle number $\mathcal{N}_2=\mathcal{N}_3$,
\item[\emph{(3)}] particle injection at constant temperature $T$ and constant volume $V_3=V_4$,
\item[\emph{(4)}] expansion at constant temperature $T$ and constant particle number $\mathcal{N}_4=\mathcal{N}_1$.
\end{itemize}
This cycle has the advantage to fix extensive quantitites, $V_j$ and $\mathcal{N}_j$ (see figure \ref{otto.graph}), that are easily controllable in some implementations.

\begin{figure*}[htbp]
\centering
\includegraphics[width=0.45\textwidth]{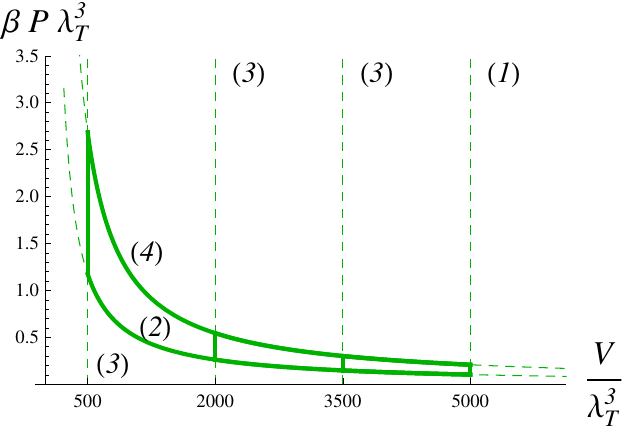}
\hspace{25pt}
\includegraphics[width=0.45\textwidth]{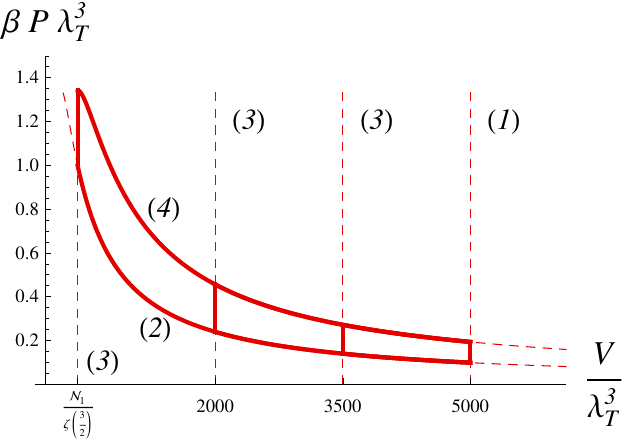}
\caption{
Rescaled pressure-volume diagram of the isothermal chemical Otto cycle for fermions (left panel, green) with $\mathcal{N}_1=1000$, $\mathcal{N}_3=500$, $V_1/\lambda_T^3=5000$ and $V_3/\lambda_T^3=500,2000,3500$, and for bosons (right panel, red) with $\mathcal{N}_1=1000$, $\mathcal{N}_3=500$, $V_1/\lambda_T^3=5000$ and $V_3/\lambda_T^3=\mathcal{N}_1/\lambda_T^3,2000,3500$. The value $V_3/\lambda_T^3=\mathcal{N}_1/\lambda_T^3$ guarantees that bosonic subtances are in a BEC phase at the end of the third stroke.
The numbers in parentheses indicate the strokes of the cycle corresponding to the closest curve.
}
\label{PV.otto}
\end{figure*}

Thermal quantities for ideal homogeneous gases are detailed in appendix \ref{app.ideal}.
Using them in
the general form of chemical work exchanged during the isothermal-isochoric strokes, shown in appendix \ref{app.transf}, one obtains

\begin{widetext}
\begin{align}
\label{WC1.Otto}
W^{\textnormal{C}}_1= & \pm\frac{V_1}{\beta\lambda_T^3}
\left(
\textnormal{Li}_{\frac{5}{2}}\textnormal{Li}^{-1}_{\frac{3}{2}}(\pm\lambda_T^3\rho_2)
-\textnormal{Li}_{\frac{5}{2}}\textnormal{Li}^{-1}_{\frac{3}{2}}(\pm\lambda_T^3\rho_1)
-\lambda_T^3\rho_2\ln\left(\pm\textnormal{Li}^{-1}_{\frac{3}{2}}(\pm\lambda_T^3\rho_2)\right)
+\lambda_T^3\rho_1\ln\left(\pm\textnormal{Li}^{-1}_{\frac{3}{2}}(\pm\lambda_T^3\rho_1)\right)
\right)\ , \\
\label{WC3.Otto}
W^{\textnormal{C}}_3= & \pm\frac{V_3}{\beta\lambda_T^3}
\left(
\textnormal{Li}_{\frac{5}{2}}\textnormal{Li}^{-1}_{\frac{3}{2}}(\pm\lambda_T^3\rho_4)
-\textnormal{Li}_{\frac{5}{2}}\textnormal{Li}^{-1}_{\frac{3}{2}}(\pm\lambda_T^3\rho_3)
-\lambda_T^3\rho_4\ln\left(\pm\textnormal{Li}^{-1}_{\frac{3}{2}}(\pm\lambda_T^3\rho_4)\right)
+\lambda_T^3\rho_3\ln\left(\pm\textnormal{Li}^{-1}_{\frac{3}{2}}(\pm\lambda_T^3\rho_3)\right)
\right)\ ,
\end{align}
\end{widetext}
where the upper (lower) signs refer to the bosonic gas without BEC (fermionic gas), $\textnormal{Li}_{s}(z)$ is the polylogarithm function \cite{Wood1992}, and $\textnormal{Li}^{-1}_{s}(z)$ its inverse ($\textnormal{Li}_{s}\textnormal{Li}^{-1}_{s}(z)=z$).
The chemical work in equations \eqref{WC1.Otto} and \eqref{WC3.Otto}, and thus the efficiency \eqref{eff.rev}
and the load per volume \eqref{load.rev}
depend only on $\lambda_T^3\rho_3$, $\lambda_T^3\rho_4$, and $v=V_3/V_1=\rho_1/\rho_4=\rho_2/\rho_3<1$.

\begin{figure*}[htbp]
\centering
\includegraphics[width=0.45\textwidth]{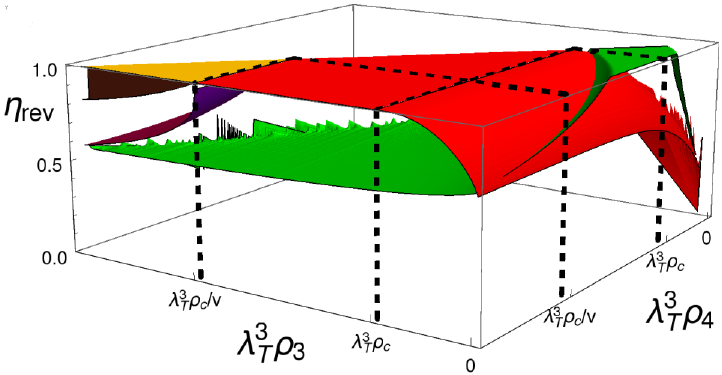}
\hspace{30pt}
\includegraphics[width=0.45\textwidth]{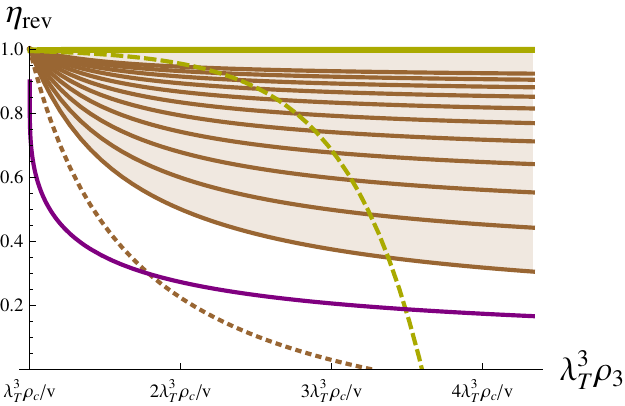}
\caption{
Efficiency $\eta_{\textnormal{rev}}$ of the isothermal chemical Otto cycle with $v=1/3$ and for ideal homogeneous gases.
Left panel: fermions (green), bosons without BECs ($\rho_j<\rho_c$) and with a BEC during the third stroke ($\rho_{1,2}<\rho_c\leqslant\rho_{3,4}$) (red), bosons in $d$D-BEC during the whole cycle ($\rho_j\geqslant\rho_c$) with \emph{(i)} $d=2$, $L_{z,3}/\lambda_T\geqslant 1$, $L_{x,3}L_{y,3}/(L_{x,1}L_{y,1})=v^{r'}$ and $r'\geqslant0$ (yellow), \emph{(ii)} $d=1$, $L_{x,3}/L_{x,1}=v^r$ and $r\geqslant0$ (brown), and \emph{(iii)} $d=0$, i.e., standard gound state BECs (purple).
Right panel: section of the left panel for $\rho_4=(1+4/v)\rho_c$ and with a $d$D-BEC during the whole cycle. The brown curves are the efficiencies for $d=1$, $L_{x,3}/L_{x,1}=v^r$ and $r$ ranging from $0$ (lower curve) to $1$ (upper curve) with step $0.1$; the brown shadow above the upper brown curve corresponds to values $r>1$.
For completeness, also the efficiency for the 2D-BEC with $L_{z,3}/\lambda_T^2=1$ and $r'=-0.2$ (yellow, dashed) and that for the 1D-BEC with $r=-0.2$ (brown, dotted) are plotted.
}
\label{otto.fig}
\end{figure*}

The load is the area within the closed line in the pressure-volume diagram plotted in figure \ref{PV.otto} for fermions (left panel, green) and bosons (right panel, red). As happens for the chemical Carnot cycle, the load is extensive and increases when the deep quantum regime is approached during the third stroke. The actual maximum is achieved at small $\rho_3$ and large $\rho_4$, that correspond to the third stroke curve in figure \ref{PV.otto} (the leftmost continuous vertical line) ranging from small to high pressures.
The efficiency $\eta_{\textnormal{rev}}$ is plotted in figure \ref{otto.fig} with $v=1/3$ for fermions (green) and bosons (red, yellow, brown, and purple). The plots for different values of $v$ are qualitatively similar.

The classical limit,
$z_j=e^{\beta\mu_j}=\lambda_T^3\rho_j\propto\epsilon\ll1$, implies vanishingly small efficiency $\eta_{\textnormal{rev}}\propto-1/\ln\epsilon$.
In the quantum regime, the efficiency $\eta_{\textnormal{rev}}$ assumes all values in the interval $[0,1]$ at different densities.
For the fermionic gas, $\eta_{\textnormal{rev}}$ has a platueax at $1$ when the signs of chemical potentials allow for $W^{\textnormal{C}}_{1,3}<0$.

The chemical potential of the bosonic gas is always negative, then $W^{\textnormal{C}}_1=-W^{\textnormal{C}}_{\textnormal{in}}$, $W^{\textnormal{C}}_3=W^{\textnormal{C}}_{\textnormal{out}}$ (with and without BECs).
If the system is a BEC during the third stroke ($\lambda_T^3\rho_{3,4}\geqslant\lambda_T^3\rho_c=\zeta(3/2)$) but not during the first stroke ($\lambda_T^3\rho_{1,2}=v\lambda_T^3\rho_{4,3}<\lambda_T^3\rho_c=\zeta(3/2)$), then
$W^{\textnormal{C}}_{\textnormal{in}}\neq0$ and $W^{\textnormal{C}}_{\textnormal{out}}\simeq0$ because $\mu_{3,4}$ approach zero and $P_3\simeq P_4$
(see appendix \ref{app.ideal}). Therefore, the efficiency is maximised, $\eta_{\textnormal{rev}}\simeq1$.
Within this parameter region, the maximum load per volume is attained when $\rho_3=\rho_c$.

\begin{figure*}[htbp]
\centering
\includegraphics[width=0.32\textwidth]{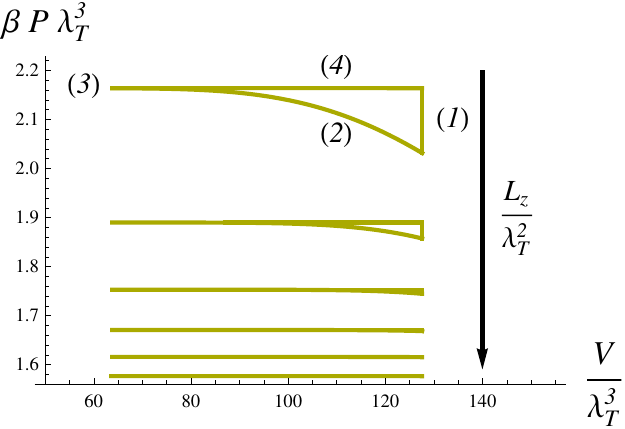}
\hspace{4pt}
\includegraphics[width=0.32\textwidth]{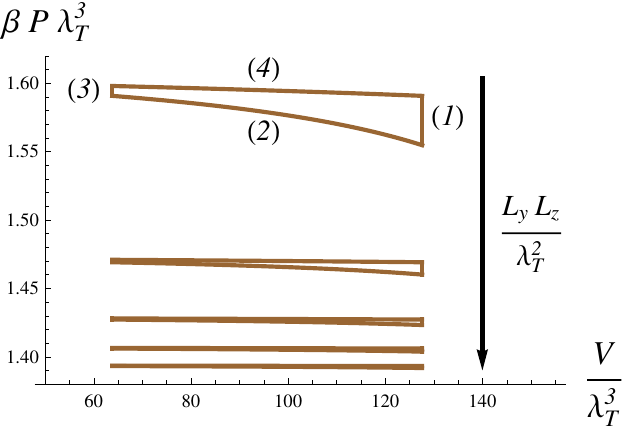}
\hspace{4pt}
\includegraphics[width=0.32\textwidth]{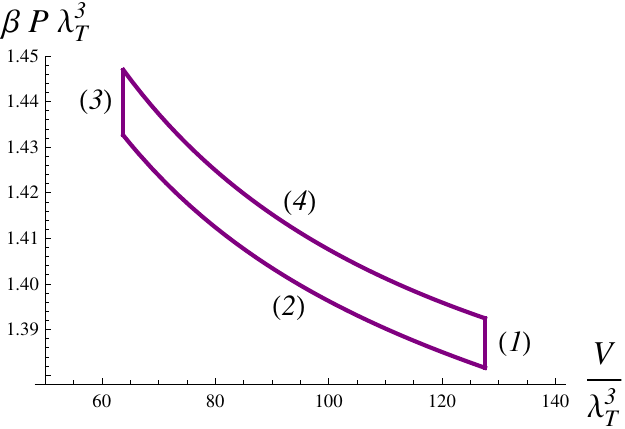}
\caption{
Rescaled pressure-volume diagram of the isothermal chemical Otto cycle when the substance is always in a BEC with $\mathcal{N}_1=1000$ and $\mathcal{N}_3=1.5 V_1\rho_c=3V_3\rho_c=500$:
2D-BEC (left panel, yellow) with increasing $L_z/\lambda_T=2,3,4,5,6,7$ in the direction of the arrow,
1D-BEC (middle panel, brown) with increasing $L_yL_z/\lambda_T^2=10,20,30,40,50$ in the direction of the arrow,
and standard ground state BEC (right panel, purple).
The numbers in parentheses indicate the strokes of the cycle corresponding to the closest curve.
}
\label{PV.otto.BEC}
\end{figure*}

If the system is a BEC also during the first stroke ($\lambda_T^3\rho_{1,2}=v\lambda_T^3\rho_{4,3}\geqslant\lambda_T^3\rho_c=\zeta(3/2)$), one derives $W^{\textnormal{C}}_{1,3}\simeq0$ from the thermodynamic quantities of different BEC phases reported in appendix \ref{app.ideal}.
The load is subextensive as shown in the pressure-volume diagram (see figure \ref{PV.otto.BEC}) for substances always in a 2D-BEC (left panel, yellow), in a 1D-BEC (middle panel, brown), and in a 0D-BEC (right panel, purple). Nevertheless, the efficiency plotted in figure \ref{otto.fig} for different types of BECs can achieve large values.

For the ground state BEC (0D-BEC), namely $L_x\sim L_y\sim L_z$, the chemical work is

\begin{align}
\label{WC1.0D-BEC}
W^{\textnormal{C}}_{1,\textnormal{0D-BEC}}= & \frac{1}{\beta}\ln\frac{v\rho_3-\rho_c}{v\rho_4-\rho_c}\ , \\
\label{WC3.0D-BEC}
W^{\textnormal{C}}_{3,\textnormal{0D-BEC}}= & \frac{1}{\beta}\ln\frac{\rho_4-\rho_c}{\rho_3-\rho_c}\ ,
\end{align}
and the efficiency $\eta_{\textnormal{rev}}$ is plotted in purple in figure \ref{otto.fig}. Note that, if $\rho_j\gg\rho_c$ for all $j=1,2,3,4$, then $\eta_{\textnormal{rev}}\to0$.

For the one-dimensional BEC (1D-BEC), namely $L_x\gtrsim\alpha' L_y L_z$,

\begin{align}
\label{WC1.1D-BEC}
W^{\textnormal{C}}_{1,\textnormal{1D-BEC}}= & \frac{\pi L_{x,1}^2}{\beta\lambda_T^2 V_1}\cdot\frac{\rho_2-\rho_1}{(\rho_2-\rho_c)(\rho_1-\rho_c)}\ , \\
\label{WC3.1D-BEC}
W^{\textnormal{C}}_{3,\textnormal{1D-BEC}}= & \frac{\pi L_{x,3}^2}{\beta\lambda_T^2 V_3}\cdot\frac{\rho_4-\rho_3}{(\rho_3-\rho_c)(\rho_4-\rho_c)}\ .
\end{align}
The corresponding efficiency $\eta_{\textnormal{rev}}$ depends on the ratio between the 3D volumes $V_3/V_1=v$ and on the ratio between the 1D volumes $L_{x,3}/L_{x,1}=v^r$. The brown regions in figure \ref{otto.fig} represent $\eta_{\textnormal{rev}}$ for $r\geqslant0$: within this interval, the efficiency increases with $r$, approaching $1$ very fast for $r>1$. When $r<0$ (brown, dotted curve in the right panel of figure \ref{otto.fig}), $\eta_{\textnormal{rev}}$ decreases to zero and then assumes negative values implying negative total mechanical work. These behaviours are more explicit in the asymptotic regime $\rho_j\gg\rho_c$ for all $j=1,2,3,4$, where $\eta_{\textnormal{rev}}\to 1-v^{2r}$.

For the two-dimensional BEC (2D-BEC), namely $L_y\gtrsim e^{\alpha L_z}\textnormal{poly}(L_z)$, the chemical work is

\begin{align}
\label{WC1.2D-BEC}
W^{\textnormal{C}}_{1,\textnormal{2D-BEC}} & =\frac{V_1 \, e^{\rho_c\lambda_T^2 L_{z,1}}}{\beta\lambda_T^2 L_{z,1}}
\left(
e^{-\rho_1\lambda_T^2 L_{z,1}}-e^{-\rho_2\lambda_T^2 L_{z,1}}\right)\ , \\
\label{WC3.2D-BEC}
W^{\textnormal{C}}_{3,\textnormal{2D-BEC}} & =\frac{V_3 \, e^{\rho_c\lambda_T^2 L_{z,3}}}{\beta\lambda_T^2 L_{z,3}}
\left(e^{-\rho_3\lambda_T^2 L_{z,3}}-e^{-\rho_4\lambda_T^2 L_{z,3}}\right)\ .
\end{align}
The efficiency as a function of $\lambda_T^3\rho_{3,4}$, plotted in yellow in figure \ref{otto.fig}, depends on the ratio between the 3D volumes $V_3/V_1=v$, on the ratio between the 2D volumes $L_{x,3}L_{y,3}/(L_{x,1}L_{y,1})=v^{r'}$ (such that $L_{z,3}/L_{z,1}=v^{1-r'}$), and on the parameter $L_{z,3}/\lambda_T$. $\eta_{\textnormal{rev}}$ is exponentially close to $1$ for the physically relevant condition $L_{z,3}/\lambda_T\geqslant1$ and for $r'\geqslant0$ (yellow, continuous curve in figure \ref{otto.fig}), while it decreases to zero and becomes negative for $r'<0$ (yellow, dashed curve in figure \ref{otto.fig}).
If $\rho_j\gg\rho_c$ for all $j=1,2,3,4$ then $\eta_{\textnormal{rev}}\simeq 1-v^{r'}e^{(v^{r'}-1)\rho_3\lambda_T^2 L_{z,3}}$ if $r'\neq0$ and $\eta_{\textnormal{rev}}\simeq 1-e^{(1-1/v)\rho_c\lambda_T^2 L_{z,3}}$ if $r'=0$.

Also in the isothermal chemical Otto cycle,
$\eta_{\textnormal{rev}}\to1$
if the substance is in a $d$D-BEC during the first stroke and in a $d'$D-BEC in the third stroke with $d'<d$, provided
$L_x\gg\alpha' L_y L_z$ for the 1D-BEC, and $L_y\gg e^{\alpha L_z}\textnormal{poly}(L_z)$ for the 2D-BEC.

In conclusion, if the substance is a BEC only during the third stroke the load $W^{\textnormal{M}}=-W^{\textnormal{C}}$ is extensive and the efficiency  is maximum $\eta_{\textnormal{rev}}\simeq 1$, while the classical limit entails poor performances. If the system remains in a BEC phase also during the first stroke, the load is subextensive, but the engine works at maximum efficiency for generalised BECs.

\section{Irreversible cycles} \label{irr}

Quasistatic transformations require infinite time and vanishing output power. Therefore, realistic engines consist in irreversible transformations implemented in finite time, $\tau_j$ for the $j$-th transformation ($j=1,2,3,4$), which provide finite output power $\pi=W^{\textnormal{M}}/\sum_j\tau_j$.
Work and heat definitions for systems exchanging energy and particles due to irreversible transformations are detailed in appendix \ref{app.transf}.
This section is devoted to show that working substances in the quantum regime provide higher power and higher efficiency at maximum power than in the classical limit.

Irreversible transformations are described by the dynamics of open systems in contact with a grandcanonical bath.
During the substance-bath interaction, the chemical potential $\mu(t)$ and the system volume $V(t)$ (and $L_x(t)$, $L_y(t)$, $L_z(t)$) slowly change in time.
For isothermal cycles, as those considered in section \ref{engines}, the bath temperature $\beta$ is constant.
The time-evolution is the solution of a time-dependent master equation with instantaneous relaxation times, $\theta(t)$, that are the relaxation times as if the explicit dependence on $t$ (through $\mu(t)$ and $V(t)$) were frozen.
Define also the maximum relaxation time $\bar\theta=\max_t\theta(t)$.
Following \cite{Cavina2017}, irreversible effects are treated perturbatively when
the dynamics is slow compared to the instantaneous relaxation, namely $\tau_j\gg\bar\theta$.
Within this regime, the limit of equilibrium quasistatic transformations is recovered
for infinite times $\tau_j/\bar\theta\to\infty$ when grandcanonical states with the same temperature and chemical potential of the bath are unique steady states at each time.
This condition
is met in master equations derived within the standard weak coupling regime (i.e. Born, Markov, and secular approximations) \cite{Schaller2011}, as in the concrete model discussed in the following.

The energy exchanges during irreversible transformations in the above perturbative regime are

\begin{align}
\label{W.irr}
W^{\textnormal{M/C}}_j & =W^{\textnormal{M/C}}_{j,\textnormal{rev}}+\frac{\bar\theta}{\tau_j}\,W^{\textnormal{M/C}}_{j,\textnormal{irr}}\ , \\
\label{Q.irr}
Q_j & =Q_{j,\textnormal{rev}}+\frac{\bar\theta}{\tau_j}\,Q_{j,\textnormal{irr}}\ , \\
\label{U.irr}
\Delta U_j & =\Delta U_{j,\textnormal{rev}}+\frac{\bar\theta}{\tau_j}\,\Delta U_{j,\textnormal{irr}}\ ,
\end{align}
where the subscript ``$\textnormal{rev}$'' denotes energy exchanges of reversible transformations.
The dependence on the initial state is an exponential decay in $\tau_j/\bar\theta$ and therefore contributes to much higher orders than those in equations \eqref{W.irr}, \eqref{Q.irr}, and \eqref{U.irr}.
In other terms, the final state of each irreversible process at the first order in $\bar\theta/\tau_j$ depends only on the initial state of the corresponding quasistatic transformation.
Consequently, the internal energy variation may not vanish after the first irreversible cycle, but the initial and the final states of all the subsequent cycles coincide
without further internal energy variations. Therefore, the internal energy variation $\Delta U$ after many cycles is negligible with respect to the total mechanical work.

The output power can be rewritten as

\begin{equation}
\pi=\frac{1}{\sum_j\tau_j}\left(W^{\textnormal{M}}_{\textnormal{rev}}+\sum_j\frac{Q_{j,\textnormal{irr}}-W^{\textnormal{C}}_{j,\textnormal{irr}}}{\tau_j}\right)\ ,
\end{equation}
where $W^{\textnormal{M}}_{\textnormal{rev}}$ is the total mechanical work of the reversible cycle.
$\pi$ is maximised at times

\begin{equation} \label{tau.max}
\tau_j^*=\frac{2\,\bar\theta\,\sqrt{W^{\textnormal{C}}_{j,\textnormal{irr}}-Q_{j,\textnormal{irr}}}}{W^{\textnormal{M}}_{\textnormal{rev}}}
\,
\sum_k\sqrt{W^{\textnormal{C}}_{k,\textnormal{irr}}-Q_{k,\textnormal{irr}}}\ ,
\end{equation}
where the physically relevant condition $W^{\textnormal{C}}_{j,\textnormal{irr}}-Q_{j,\textnormal{irr}}\geqslant 0$ has been considered. This condition implies that the irreversibility contributions decrease the internal energy and that the optimal times $\tau_j^*$ are finite.

The thermochemical engine then operates with the optimal times $\tau_j^*$ \eqref{tau.max} for each stroke, unless they approach from above the relaxation time $\bar\theta$.
In this case, the long-time perturbative approach loses its validity and the engine working times are set at $\tau_j^*=\bar\theta/s$ with $s\ll1$. In other words, the parameter $s$ has the role to move the times $\tau_j^*$ as close as possible to the boundary of perturbative regime where $\pi_*$ is maximised.
Using these optimal times, define the maximum power $\pi_*$
and the efficiency at maximum power $\eta_*$.

As a concrete model, consider that the substance is an ideal bosonic homogeneous gas
and the bath is made of harmonic oscillators, with Hamiltonian
$H_B=\hslash\int\textnormal{d}\omega\,\omega\,b_{\omega}^\dag b_{\omega}$,
and canonical commutation relations
$[b_{\omega},b_{\omega'}^\dag]=\delta(\omega-\omega')$.
The substance-bath interaction Hamiltonian is $H_I=\lambda\sum_{\bm p}\int\textnormal{d}{\bm \omega}\,h(\hslash\omega_x+\hslash\omega_y+\hslash\omega_z)\,T_{{\bm\omega},{\bm p}}\big(a_{\bm p}^\dag b_{\bm\omega}+b_{\bm\omega}^\dag a_{\bm p}\big)$ with coefficients $T_{{\bm\omega},{\bm p}}$ discussed in appendix \ref{app.irr}.
The master equation of the substance dynamics in the weak coupling regime and its long time dynamics are also reported in appendix \ref{app.irr}. In particular, instantaneous relaxation times scale as $\bar\theta=\mathcal{O}(L_z/V)$, where $L_x$, $L_y$, $L_z$, and $V$ denote typical values during the cycle here and in the following size scalings.

The first order corrections of work \eqref{W.irr} and heat \eqref{Q.irr}
are explicitly written in equations \eqref{WM.irr}, \eqref{WC.irr}, and \eqref{QQ.irr},
whose size scalings are estimated in appendix \ref{en.corr}.
These estimations are used to derive the size scalings of the optimal times $\tau_j^*$, of the maximum power $\pi_*$ and of the corresponding efficiency $\eta_*$ (see appendix \ref{app.cycles}).

\begin{figure*}[htbp]
\centering
\includegraphics[width=0.45\textwidth]{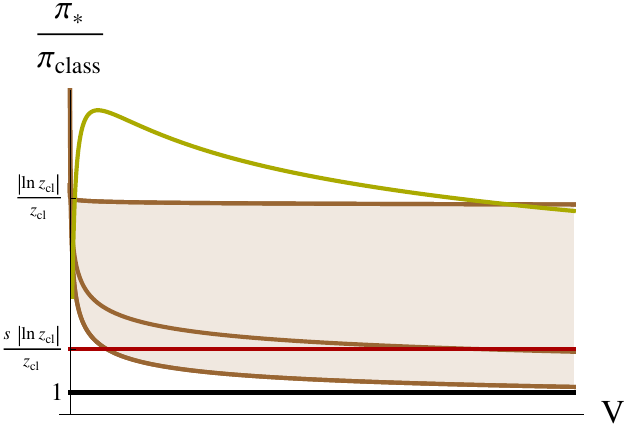}
\hspace{30pt}
\includegraphics[width=0.45\textwidth]{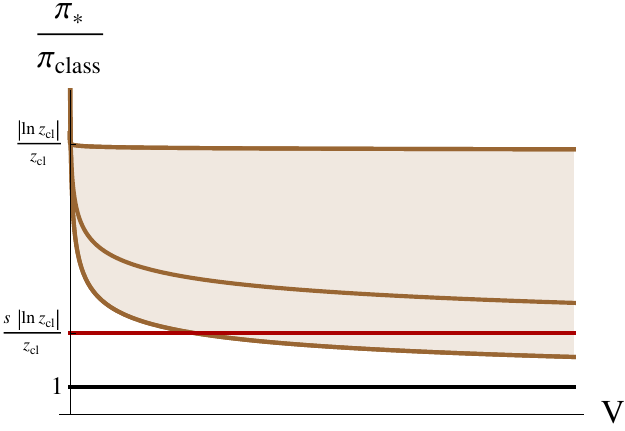}
\caption{
Size scaling of the maximum power normalised to its classical limit.
Left panel: irreversible isothermal chemical Carnot cycle with 2D-BEC (yellow), with 1D-BEC (brown) for $\chi=1.01,1.5,2$ from the upper to the lower curve, with 0D-BEC or without BECs (red).
Right panel: irreversible isothermal chemical Otto cycle with 1D-BEC (brown) for $\chi=2.01,2.5,3$ from the upper to the lower curve, with other BECs or without BECs (red).
The black line is the classical limit, and $s>z_{\textnormal{cl}}/|\ln z_{\textnormal{cl}}|$ has been considered.
}
\label{power.graph}
\end{figure*}

In the following subsections, the performances $\pi_*$ and $\eta_*$ of irreversible chemical Carnot and Otto cycles in quantum regimes are shown to outperform those in the classical limit.
The size scalings of the optimal power normalised to its classical limit, $\pi_*/\pi_{\textnormal{class}}$ are summarised in figure \ref{power.graph} for Carnot (left panel) and Otto (right panel) cycles
when the substance during the third stroke is a 2D-BEC (yellow for the Carnot cycle and red for the Otto cycle), in a 1D-BEC (brown), a ground state BEC or without BECs (red), and in the classical limit (black). Note that the values of $\pi_*/\pi_{\textnormal{class}}$ in the quantum regime, as well as
the accessible area for 1D-BECs (brown), are very large
since the fugacity in the classical limit is $z_{\textnormal{cl}}\ll1$.

\subsection{Irreversible cycles without BECs and in the classical limit}

In the absence of BEC phases during the cycle, the computations in appendices \ref{en.corr} and \ref{app.cycles} prove that the maximum power scales as $\pi_*=\mathcal{O}\big(\frac{s\,V^2}{L_z}\big)$ and the efficiency as $\eta_*=\eta_{\textnormal{rev}}+\mathcal{O}(s\,V^0)$.
The factor $s\ll1$ has been introduced in order to remain in the perturbative regime, when the optimal times in equation \eqref{tau.max} scale as the relaxation times $\bar\theta=\mathcal{O}(L_z/V)$, as discussed above.

In the classical limit, the condition \eqref{class}, that reads $\mathcal{N}\simeq z_{\textnormal{cl}}V/\lambda_T^3$ with small fugacity $z_{\textnormal{cl}}=e^{\beta\mu_{\textnormal{cl}}}\ll1$ for ideal gases (see appendices \ref{app.class} and \ref{app.ideal}), affects the above scalings.
In particular, the optimal time from equation \eqref{tau.max} is $\tau_{\textnormal{class}}=\mathcal{O}\big(\frac{L_z}{V}|\ln z_{\textnormal{cl}}|\big)$, the maximum power is $\pi_{\textnormal{class}}=\mathcal{O}\big(\frac{z_{\textnormal{cl}}\,V^2}{|\ln z_{\textnormal{cl}}|L_z}\big)$, and the efficiency at maximum power
reads $\eta_{\textnormal{class}}=\mathcal{O}(z_{\textnormal{cl}})$.

In the cycles discussed in section \ref{engines}, $\eta_{\textnormal{rev}}$ is finite and approaches one close to BEC transitions, and thus the quantum regime even without BECs exhibits much higher efficiency than the classical limit and larger power if $s>z_{\textnormal{cl}}/|\ln z_{\textnormal{cl}}|$.

\subsection{Irreversible cycles with BECs} \label{irr.cycles.BEC}

When the working substance of the engine undergoes a BEC transition, one has to carefully consider the confinement anisotropy that leads to standard or generalised BECs, in order to rephrase the size scalings in terms of the volume.
In the following, three different anisotropic boxes are considered, each one favouring a different BEC phase.
The sizes $L_z=\mathcal{O}(\ln V)$ and $L_x\simeq L_y=\mathcal{O}\big(\sqrt{V/\ln V}\big)$ favours the formation of a 2D-BEC.
The emergence of a 1D-BEC is studied when $L_x=\mathcal{O}(L_y L_z)^\chi$ with $\chi\geqslant 1$, such that $L_x=\mathcal{O}\big(V^{\frac{\chi}{\chi+1}}\big)$ and $L_y\sim L_z=\mathcal{O}\big(V^{\frac{1}{2\chi+2}}\big)$.
Lastly, assume the scaling $L_x\sim L_y\sim L_z=\mathcal{O}\big(\sqrt[3]{V}\big)$ when the substance is a 0D-BEC.

The size scaling of the work and heat corrections in equations \eqref{W.irr} and \eqref{Q.irr} depend on the function $h(\varepsilon)$, and are explicitly computed in appendix \ref{en.corr} assuming $h(\varepsilon_p)\neq0$ and that $h(0)$ is finite. The condition $h(\varepsilon_p)\neq0$ guarantees that the grandcanonical ensemble is the unique steady state at every time, otherwise also the instantaneous Hamiltonian eigenstates with energy $\varepsilon_p$ such that $h(\varepsilon_p)=0$ are steady states. This large class of functions contains the constant function, the exponential decay, and the Lorentzian.

The performances of the isothermal chemical Carnot and Otto cycles are reported explicitly in the following, when the substance is a BEC only during the third stroke. These configurations show efficiency and output power overcoming the classical ones.
The cases of the working substance always being in a BEC are discussed in appendix \ref{app.cycles}.

\subsubsection{Irreversible chemical Carnot cycle with a BEC during the third stroke} \label{irr.Carnot3}

The size scalings of irreversible energy corrections imply the following efficiency at maximum power for the isothermal chemical Carnot cycle with a BEC only during particle injection
(see appendix \ref{app.cycles})

\begin{equation}
\label{eta.max.BEC}
\eta_*=
\begin{cases}
\displaystyle
\frac{\eta_{\textnormal{rev}}}{2}
+\mathcal{O}\left(\frac{\big(\ln V\big)^5}{V}\right)^\frac{1}{4}
& \displaystyle \textnormal{2D-BEC} \\
\displaystyle
\frac{\eta_{\textnormal{rev}}}{2}
+\mathcal{O}\left(\frac{1}{V^\frac{\chi-1}{2\chi+2}}\right)
& \displaystyle \textnormal{1D-BEC, } \chi>1 \\
\displaystyle
\eta_{\textnormal{rev}}
+\mathcal{O}\left(s\,V^0\right)
& \displaystyle \textnormal{other BECs}
\end{cases}\ .
\end{equation}
Recall that $\eta_{\textnormal{rev}}=1-\mu_3/\mu_1\simeq 1$ for large size, as discussed in section \ref{Carnot}, and finite size corrections (through those of $\mu_3$ in the BEC phases) are smaller that those due to irreversibility in equation \eqref{eta.max.BEC}.
Therefore, the efficiencies \eqref{eta.max.BEC} are much larger than in the classical limit $\eta_{\textnormal{class}}=\mathcal{O}(z_{\textnormal{cl}})\ll1$, where $z_{\textnormal{cl}}$ is the classical fugacity.

The ratio between the maximum power in BEC phases and the maximum power in the classical limit with the same $L_{x,y,z}$ is

\begin{equation}
\frac{\pi_*}{\pi_{\textnormal{class}}}=
\begin{cases}
\displaystyle
\mathcal{O}\left(
\frac{|\ln z_{\textnormal{cl}}|}{z_{\textnormal{cl}}} \, \frac{\big(\ln V\big)^\frac{5}{2}}{\sqrt{V}}\right)
& \displaystyle \textnormal{2D-BEC} \\
\displaystyle
\mathcal{O}\left(
\frac{|\ln z_{\textnormal{cl}}|}{z_{\textnormal{cl}}\,V^{\frac{2\chi-2}{2\chi+2}}}
\right)
& \displaystyle \textnormal{1D-BEC, } \chi>1 \\
\displaystyle
\mathcal{O}\left(
\frac{s\,|\ln z_{\textnormal{cl}}|}{z_{\textnormal{cl}}}
\right)
& \displaystyle \textnormal{other BECs}
\end{cases}\ .
\end{equation}
The output power is larger than the classical limit
if $V/(\ln V)^5<|\ln z_{\textnormal{cl}}|^2/z_{\textnormal{cl}}^2$ for 2D-BECs,
if $V<\big(|\ln z_{\textnormal{cl}}|/z_{\textnormal{cl}}\big)^{\frac{2\chi+2}{2\chi-2}}$ for 1D-BECs with $\chi>1$, and when $s>z_{\textnormal{cl}}/|\ln z_{\textnormal{cl}}|$ for 0D-BECs and 1D-BECs with $\chi=1$.

\subsubsection{Irreversible chemical Otto cycle with a BEC during the third stroke} \label{irr.Otto3}

The size scalings for the irreversible isothermal chemical Otto cycle, with BECs only when chemical work is released, provide the following efficiency at maximum power (see appendix \ref{app.cycles})

\begin{equation}
\label{eta.max.BECOtto}
\eta_*=
\begin{cases}
\displaystyle
\frac{\eta_{\textnormal{rev}}}{2}+
\mathcal{O}\left(\frac{1}{V^\frac{\chi-2}{2\chi+2}}\right)
& \displaystyle \textnormal{1D-BEC, } \chi>2 \\
\displaystyle
\eta_{\textnormal{rev}}+
\mathcal{O}\left(s\,V^0\right)
& \displaystyle \textnormal{other BECs}
\end{cases}\ .
\end{equation}
From figure \ref{otto.fig}, $\eta_{\textnormal{rev}}\simeq 1$, so that the efficiencies \eqref{eta.max.BECOtto} are much larger than in the classical limit $\eta_{\textnormal{class}}=\mathcal{O}(z_{\textnormal{cl}})\ll1$, where $z_{\textnormal{cl}}$ is the classical fugacity.

The maximum power in BEC phases normalised to that in the classical limit with the same $L_{x,y,z}$ is

\begin{equation}
\frac{\pi_*}{\pi_{\textnormal{class}}}=
\begin{cases}
\displaystyle
\mathcal{O}\left(
\frac{|\ln z_{\textnormal{cl}}|}{z_{\textnormal{cl}}\,V^{\frac{\chi-2}{\chi+1}}}
\right)
& \displaystyle \textnormal{1D-BEC, } \chi>2 \\
\displaystyle
\mathcal{O}\left(
\frac{s\,|\ln z_{\textnormal{cl}}|}{z_{\textnormal{cl}}}
\right)
& \displaystyle \textnormal{other BECs}
\end{cases}\ .
\end{equation}
Therefore, the output power is larger than the classical limit
if $V<\big(|\ln z_{\textnormal{cl}}|/z_{\textnormal{cl}}\big)^{\frac{\chi+1}{\chi-2}}$ for 1D-BECs with $\chi>2$, and when $s>z_{\textnormal{cl}}/|\ln z_{\textnormal{cl}}|$ for other BECs.

\section{Conclusions and discussion} \label{concl}

This paper explores the performances of
quantum thermochemical engines, that convert chemical work into mechanical work by putting a working substance in contact with thermochemical sources.
The sources control different thermodynamic quantities of the substance, allowing for several thermodynamic strokes.
For the sake of concreteness, the isothermal chemical Carnot and Otto cycles are discussed in details.

When the working substance is a quantum (either fermionic or bosonic) degenerate gas, the above cycles achieve the full range of efficiency $\eta\in[0,1]$ and approach the maximum efficiency for a wide region of system parameters. On the other hand, the classical limit implies low efficiency and small output mechanical work.
In other words, the chemical potential values that minimise the wasted chemical work leave the validity domain of the classical limit.
The maximum efficiency is achieved when the chemical work is always done on the substance without waste, and corresponds to negative (non-negative) chemical potential during particle release (injection).
Therefore, quantum statistics enhances thermochemical engine performances,
as already proven for heat engines \cite{Myers2020,Gupt2021,Myers2022,Sur2023,Yadin2023}, batteries \cite{Konar2022}, metrology \cite{Braun2018,Marzolino2013,Potts2019}, and information protocols \cite{Benatti2020,Johann2021,Benatti2021}.

This paper presents a detailed analysis
for ideal homogeneous gases in the grandcanonical ensemble.
Special attention has been devoted to the role of BEC phases where the ground state alone, or eigenstates with parallel momenta, or those with coplanar momenta are macroscopically occupied according to the confinement anisotropy. The presence of BEC phases only during the particle injection entails maximum efficiency and extensive output mechanical work for reversible bosonic cycles. If the substance is in BEC phases during the entire cycle, the efficiency is maximum for a larger parameter region but with subextensive output mechanical work.

These results are generalised in several directions.
Firstly, although the grandcanonical ensemble provides the simplest description of quantum statistics, equivalent equation of states and chemical potential ranges are found in the so-called $\mu PT$ ensemble \cite{Hill,Campa2018,Marzolino2021}, that allows also for volume statistical fluctuations and are studied for small systems and nanothermodynamics \cite{Chamberlin2000,Hill2001,Hill2002,Qian2012,Chamberlin2015,Latella2015,Bedeaux2018}. Therefore, the aforementioned quantum enhanced performances of thermochemical engines are observed also in the $\mu PT$ ensemble.

A second generalisation is the presence of general trapping potentials and density of states \cite{deGroot1950,Oliva1989,Yan1999,MartinezHerrera2019,Momeni2020}. In this cases the range of chemical potentials are not changed and the equations of state show the same qualitative behaviour of the ideal homogeneous gas, e.g., with upper-bounded bosonic particle number in the continuum approximation and the emergence of BEC phases above critical densities and at vanishing chemical potentials.
Also these generalisations allow for enhanced performances for converting chemical into mechanical work in quantum regimes.

The inclusion of interactions neither limits the accessible range of chemical potentials \cite{PethickSmith,PitaevskiiStringari,Giorgini2008},
such that enhanced performances of thermochemical engines are again observed in the quantum regime. The emergence of standard BEC have been intensively studied for interacting gases under the Bogoliubov, Hartree-Fock and Thomas-Fermi approximations. Standard as well as generalised BEC are proven also in effective mean-field models \cite{Zagrebnov2001} and van der Waals interactions (see appendix \ref{vanderWaals}) above a critical density.
The features of quantum statistics, that can be employed as a resource for thermochemical engines, then persist in several physical models.

The quasistaticity of reversible cycles implies vanishing output power, and points at the need of considering irreversible cycles.
Irreversible cycles have been modeled through interactions with a grandcanonical bath and slowly varying control parameters
for process durations much larger than instantaneous relaxation times.
The classical limit still provides small efficiency and output mechanical work, while quantum gases, especially with BECs during the particle injection, entail much larger efficiencies and much larger output power at comparable volumes.

In conclusion, the aforementioned results indicate a quantum advantage for the efficiency and the power of machines that convert chemical work into mechanical work.
Exploiting existing realisations of quantum degenerate gases \cite{PethickSmith,PitaevskiiStringari,Giorgini2008} and recent developments in quantum simulators \cite{Georgescu2014,Monroe2020},
the thermochemical engines discussed in this paper could open the way for the development of new quantum enhanced engines and motors.
Moreover, quantum advantages occur also without the need of a BEC but away from the classical limit, and could shed light on phenomena at the border between the quantum and the classical domain.

\appendix

\section{Classical limit of quantum statistical mechanics} \label{app.class}

Quantum systems behave as classical models when
fundamental aspects of quantum mechanics have negligible effects. In statistical mechanics, this condition is realised when elementary cells in the single-particle phase space, with volume $\hslash^3$, can be approximated with points \cite{Gallavotti1999}, for instance when the number of particles is much smalled than the number of elementary cells. Therefore, at most one particle lies in an elementary cell and quantum statistics due to indistinguishability reduce to Boltzmann statistics.

The number of elementary cells is the extension of the thermal state in the single-particle phase space, e.g., estimated by the variances of canonical variables, divided by $\hslash^3$.
The above condition then reads

\begin{equation} \label{class.cond}
\mathcal{N}\ll\frac{\Delta x\Delta y\Delta z\Delta p_x\Delta p_y\Delta p_z}{\hslash^3},
\end{equation}
where $(x,y,z)$ and $(p_x,p_y,p_z)$ are the position and momentum operators of just one particle, $\Delta x=\sqrt{\langle x^2\rangle-\langle x\rangle^2}$, $\Delta p_x=\sqrt{\langle p_x^2\rangle-\langle p_x\rangle^2}$, and similarly for the other variances with grandcanonical averages $\langle\cdot\rangle$. The symmetry under particle permutation in systems of identical particles implies that every particle provides the same variances in Eq.~\eqref{class.cond}.

The product $V_*=\Delta x\Delta y\Delta z$ is proportional to the volume occupied by the system: e.g., it can be easily computed that $V_*\propto L_xL_yL_z$ for the homogeneous gas,
and $V_*\propto(m^3\beta^3\omega_x^2 \omega_y^2 \omega_z^2)^{-\frac{1}{2}}$ for the ideal gas in a harmonic potential where the large volume limit is $\omega_{x,y,z}\to0$ and the density is proportional to $\mathcal{N}\omega_x \omega_y \omega_z$ \cite{Mullin1997}.
Moreover, $V_*$ increases when repulsive interactions dominate, and decreases for strong attractive interactions.
If the Hamiltonian is invariant under time reversal (${\bm p}\leftrightarrow-{\bm p}$), for instance for momentum independent interactions and external potentials, then $\langle p_{x,y,z}\rangle=0$,
and one estimates $\Delta^2 p_{x,y,z}=\langle p_{x,y,z}^2\rangle\leqslant\langle p_x^2\rangle+\langle p_y^2\rangle+\langle p_z^2\rangle\simeq 2m K/\mathcal{N}$,
where $K$ is the average kinetic energy of the system.
Therefore, the condition \eqref{class.cond} becomes

\begin{equation}
\mathcal{N}\ll V_*\left(\frac{2mK}{\hslash^2\mathcal{N}}\right)^{\frac{3}{2}}\propto V\left(\frac{2mK}{\hslash^2\mathcal{N}}\right)^{\frac{3}{2}}\ ,
\end{equation}
or, after simple manipulations,

\begin{equation}
\mathcal{N}\ll V\left(\frac{2mK}{\hslash^2 V}\right)^{\frac{3}{5}}\ ,
\end{equation}
where $K/\mathcal{N}$ and $K/V$ are intensive quantities.

The expectations $\langle p_{x,y,z}^2\rangle$ can be written in a mathematically elegant form, known as the quantum counterpart of the equipartition theorem \cite{Bialas2019,Luczka2020}, which reduces to the classical formula $\langle p_{x,y,z}^2\rangle=mk_B T$ for high temperature or for the ideal homogeneous gas. Since Eq.~\eqref{class.cond} is the condition for quantum gases behaving classically, it is meaningful to plug the classical formula $\langle p_{x,y,z}^2\rangle=mk_B T$ there, thus obtaining $\mathcal{N}\ll V/\lambda_T^3$ up to a multiplicative constant, where $\lambda_T=\sqrt{2\pi\hslash^2\beta/m}$ is thermal wavelength. This form is equivalent to small fugacities $z=e^{\beta\mu}\ll 1$ for ideal gases (see appendix \ref{app.ideal}).

The constraint implied by the classical limit impinges on the work production of thermochemical cycles.
Indeed, the chemical work during each stroke is estimated using the integral mean value theorem: denoting the integral along the $j$-th stroke by $\int_j$, and reminding that $\mathcal{N}_j$ is the particle number at the beginning of the $j$-th stroke (with $\mathcal{N}_{J+1}=\mathcal{N}_1$ for a cycle consisting of $J$ strokes),
one obtains

\begin{align}
\left|W^{\textnormal{C}}_j\right|= & \left|\int_j \mu\,\textnormal{d}\mathcal{N}\right|=\Big|\bar{\mu}_j\big(\mathcal{N}_{j+1}-\mathcal{N}_j\big)\Big| \nonumber \\
\ll & \left|\bar{\mu}_j\right|\mathcal{O}\big(\max\{V_j,V_{j+1}\}\big)\ ,
\end{align}
where $\bar{\mu}_j$ is an average chemical potential during the $j$-th stroke. For reversible isothermal thermochemical cycles, the load, namely the mechanical work, is then $W^{\textnormal{M}}=-W^{\textnormal{C}}\ll\mathcal{O}(\max V)$ where $\max V$ is the maximum volume attained during the cycle.

\section{Ideal homogeneous gas} \label{app.ideal}

An ideal homogeneous gas consists in non-interacting particles confined in a cube of sizes $L_{x,y,z}$ \cite{Reichl2016}.
The gas Hamiltonian is $H=\sum_{\bm p}\varepsilon_p a_{\bm p}^\dag a_{\bm p}$ with $\varepsilon_p=(p_x^2+p_y^2+p_z^2)/(2m)=p^2/(2m)$, and the momenta ${\bm p}=(p_x,p_y,p_z)=2\pi\hslash\,(n_x/L_x,n_y/L_y,n_z/L_z)$ label its eigenmodes ($n_{x,y,z}\in\mathbbm{Z}$ and $[a_{\bm p},a_{{\bm p}'}^\dag]=\delta_{{\bm p},{\bm p}'}$).
Thermodynamic quantities satisfy the following relations

\begin{align}
\label{Uid}
U & = \pm\frac{3V}{2\beta\lambda_T^3}\textnormal{Li}_{\frac{5}{2}}(\pm e^{\beta\mu})=\frac{3}{2}PV\ , \\
\label{Nid}
\mathcal{N} & =\pm\frac{V}{\lambda_T^3}\textnormal{Li}_{\frac{3}{2}}(\pm e^{\beta\mu})\ ,
\end{align}
where the upper signs holds for bosonic particles and the lower sign for fermions, $\lambda_T=\sqrt{2\pi\hslash^2\beta/m}$ is the thermal wavelength, $m$ is the particle mass,

\begin{equation} \label{polylog}
\textnormal{Li}_s(z)=\sum_{k=1}^\infty\frac{z^k}{k^s}=\frac{1}{\Gamma(s)}\int_0^\infty\textnormal{d}t\frac{t^{s-1}}{z^{-1}e^t-1}
\end{equation}
is the polylogarithm function \cite{Wood1992} and $\Gamma(s)$ is the Gamma function. The following series representations \cite{Wood1992} will be used for estimating thermal quantities at small chemical potentials:

\begin{equation} \label{series.nonint}
\textnormal{Li}_s(e^{\beta\mu})=\Gamma(1-s)(-\beta\mu)^{s-1}
+\sum_{k=0}^\infty\frac{\zeta(s-k)}{k!}(\beta\mu)^k
\end{equation}
if $s\notin\mathbbm{N}^+$, and

\begin{align} \label{series.int}
\textnormal{Li}_s(e^{\beta\mu})= & \frac{(\beta\mu)^{s-1}}{(s-1)!}\big(H_{s-1}-\ln(-\beta\mu)\big) \nonumber \\
& +\sum_{\substack{k=0\\k\neq s-1}}^\infty\frac{\zeta(s-k)}{k!}(\beta\mu)^k
\end{align}
if $s\in\mathbbm{N}^+$, where $\displaystyle H_s=\sum_{n=1}^s 1/n$ is the harmonic number.

The condition for the energy eigenstate occupancies of bosonic gases $\big(e^{\beta(\varepsilon_p-\mu)}-1\big)^{-1}\geqslant0$ contraints the chemical potential to non-positive values, approaching zero at the formation of a Bose-Einstein condensate.
The chemical potential of the fermionic gas has no restrictions. The classical limit is achieved for small fugacities $z=e^{\beta\mu}\ll1$, thus at negative chemical potentials, which implies $\textnormal{Li}_s(z)\simeq z$ and the known equation of state $PV=\mathcal{N}k_BT$ from Eqs.~(\ref{Uid},\ref{Nid}).

The density $\rho=\mathcal{N}/V$ of the bosonic gas, from Eq.~\eqref{Nid}, has an upper bound when $z=1$ or $\mu=0$:

\begin{equation}
\rho=\frac{\mathcal{N}}{V}\leqslant\frac{1}{\lambda_T^3}\,\zeta\left(\frac{3}{2}\right)\equiv\rho_c\ ,
\end{equation}
also called critical density, where $\zeta(s)$
is the Riemann zeta function.
For densities larger than $\rho_c$,
$\mathcal{N}>\rho_c V$, or equivalently when the temperature is lowered below the critical one, i.e.,

\begin{equation} \label{Tc}
T_c=\frac{2\pi\hslash^2}{mk_B}\left(\frac{\rho}{\zeta\left(\frac{3}{2}\right)}\right)^{\frac{2}{3}},
\end{equation}
the chemical potential approaches zero, and the fraction of particles

\begin{equation}
f=\frac{\mathcal{N}-\rho_c V}{\mathcal{N}}=1-\left(\frac{T}{T_c}\right)^{\frac{3}{2}}
\end{equation}
accumulates in a Bose-Einstein condensate (BEC).

The nature of the BEC depends on the relative scaling of the box sizes.
The BEC consists in a macroscopic number of particles in the ground state for isotropic confinement volume \cite{PethickSmith,PitaevskiiStringari},
 but is modelled by a two- or a one-dimensional gas in highly anisotropic external potentials
\cite{Krueger1967,Krueger1968,Rehr1970,vanDenBerg1982,vanDenBerg1983,vanDenBerg1986-1,
vanDenBerg1986-2,Ketterle1996,vanDruten1997,Beau2010,Mullin2012}.
Assuming $L_x\geqslant L_y\geqslant L_z$,
there are three different scenarios.

If $L_y\gtrsim e^{\alpha L_z}\textnormal{poly}(L_z)$, where $\alpha$ is a constant and $\textnormal{poly}(L_z)$ stands for a polynomial in $L_z$, the BEC consists in the macroscopic occupation of the effective two-dimensional gas made of states with momenta perpendicular to $L_z$ (2D-BEC). For this reason, $\mathcal{O}\big(\frac{\lambda_T^2}{L_y^2}\big)\leqslant-\beta\mu\ll\frac{\lambda_T^2}{L_z^2}$. Moreover, the condensate occupation is

\begin{align}
f\,\mathcal{N} & =\frac{L_xL_y}{\lambda_T^2}\,\textnormal{Li}_{1}(e^{\beta\mu})=-\frac{L_xL_y}{\lambda_T^2}\ln\left(1-e^{\beta\mu}\right) \nonumber \\
& \underset{\mu\simeq0}{\simeq}-\frac{L_xL_y}{\lambda_T^2}\ln(-\beta\mu)\ ,
\end{align}
such that the chemical potential is $-\beta\mu\simeq e^{-f\rho L_z\lambda_T^2}$.
Using this scaling of $\mu$, the total pressure of the non-condensed part of the gas and of the 2D-BEC is

\begin{align}
P & =\frac{\textnormal{Li}_{\frac{5}{2}}(e^{\beta\mu})}{\beta\lambda_T^3}+\frac{\textnormal{Li}_2(e^{\beta\mu})}{\beta\lambda_T^2L_z} \nonumber \\
& \underset{\mu\simeq0}{\simeq}\frac{\zeta(\frac{5}{2})}{\beta\lambda_T^3}+\frac{\zeta(2)}{\beta\lambda_T^2L_z}
-\frac{\mu\ln(-\beta\mu)}{\lambda_T^2 L_z}
+\frac{\mu}{\lambda_T^2L_z}\ .
\end{align}

If $L_x\gtrsim\alpha' L_y L_z$, for a constant $\alpha'$, the BEC is formed by an effective one-dimensional gas consisting of states with momenta parallel to $L_x$ (1D-BEC). Therefore, $\mathcal{O}\big(\frac{\lambda_T^2}{L_x^2}\big)\leqslant-\beta\mu\ll\frac{\lambda_T^2}{L_y^2}$. The condensate occupation is

\begin{equation}
f\,\mathcal{N}=\frac{L_x}{\lambda_T^2}\,\textnormal{Li}_{\frac{1}{2}}(e^{\beta\mu})\underset{\mu\simeq0}{\simeq}\frac{L_x}{\lambda_T^2}\sqrt{\frac{\pi}{-\beta\mu}}\ ,
\end{equation}
and consequently the chemical potential is $-\beta\mu\simeq\pi(f\rho L_y L_z\lambda_T)^{-2}$.
The correction to the pressure due to the 1D-BEC at the lowest orders is

\begin{align}
P & =\frac{\textnormal{Li}_{\frac{5}{2}}(e^{\beta\mu})}{\beta\lambda_T^3}+\frac{\textnormal{Li}_{\frac{3}{2}}(e^{\beta\mu})}{\beta\lambda_TL_yL_z} \nonumber \\
& \underset{\mu\simeq0}{\simeq}\frac{\zeta(\frac{5}{2})}{\beta\lambda_T^3}+\frac{\zeta(\frac{3}{2})}{\beta\lambda_TL_yL_z}
+\sqrt{-\frac{\mu}{\beta}}\ \frac{\Gamma(-\frac{1}{2})}{\lambda_TL_yL_z}
\end{align}

If $L_x\sim L_y\sim L_z$, a standard BEC with the ground state macroscopically occupied is created (0D-BEC), with condensate number

\begin{equation}
f\,\mathcal{N}=\frac{1}{e^{-\beta\mu}-1}\underset{\mu\simeq0}{\simeq}-\frac{1}{\beta\mu}\ ,
\end{equation}
and chemical potential $-\beta\mu=(f\rho V)^{-1}$.
The pressure of the non-condensed gas and the 0D-BEC is

\begin{equation}
P=\frac{\textnormal{Li}_{\frac{5}{2}}(e^{\beta\mu})}{\beta\lambda_T^3}+\frac{\textnormal{Li}_1(e^{\beta\mu})}{\beta L_xL_yL_z}
\underset{\mu\simeq0}{\simeq}\frac{\zeta(\frac{5}{2})}{\beta\lambda_T^3}-\frac{\ln(-\beta\mu)}{\beta V}
\end{equation}

If the volume sizes fulfil more that one among the above scalings, subsequent BEC transitions happen from a $d$D-BEC to a $d'$D-BEC with $d'<d$ when the temperature decreases or the density increases.

Quantum gases in more general trapping potentials and density of states exhibit average particle numbers similar to equation \eqref{Nid}, with the polylogarithm function $\textnormal{Li}_{\frac{3}{2}}(\pm e^{\beta\mu})$ replaced by $\textnormal{Li}_{s}(\pm e^{\beta\mu})$ and a model-dependent parameter $s$ \cite{deGroot1950,Oliva1989,Yan1999}.
The average particle number is expressed in terms of these functions also in several weak, mean-field and hard-core interactions \cite{Zagrebnov2001,Dai2007,OlivaresQuiroz2011,Vovchenko2015,Bhuiyan2021} (see for instance appendix \ref{vanderWaals}).
This mathematical analogy allows us to extend some qualitative features of the homogeneous gases to more general systems.
In particular, BECs above a critical density occur when the polylogarithm functions in the expressions of $\mathcal{N}$ are bounded from above, i.e., $s>1$ (see equation \eqref{series.nonint}. BECs above a critical density also occur with different mathematical forms of $\mathcal{N}$ \cite{MartinezHerrera2019,Momeni2020}.

\section{Van der Waals gases} \label{vanderWaals}

The classical van der Waals model describes hard-core particles interacting through the Lennard-Jones potential within a mean-field approximation \cite{Johnston,Vovchenko2015-2}. The equation of state is

\begin{equation} \label{vdW}
(P+a\rho^2)(1-b\rho)=\rho\,k_B T,
\end{equation}
where $b>0$ is the volume excluded for each particle by other hard-core particles and $a$, i.e., the average interaction per unit density, is proportional to $b$ and to a characteristic energy scale, $\phi$, of the potential. The chemical potential is

\begin{equation} \label{chem.pot.vdW}
\mu=k_B T\ln(\lambda_T^3\rho)-k_B T\ln(1-b\rho)+\frac{k_B T\,b\rho}{1-b\rho}-2a\rho\ .
\end{equation}

The first contribution in Eq.~\eqref{chem.pot.vdW} is the chemical potential of the ideal gas, $k_B T\ln(\rho\,\lambda_T^3)$, that is negative with large magnitude from the classical limit $\rho=\mathcal{N}/V\ll 1/\lambda_T^3$ derived in appendix \ref{app.class}. Moreover, $b\rho\leqslant1$ from the  definition of $b$, and the maximum density $\rho=1/b$ is approached at vanishing temperature, i.e., $1-b\rho\propto k_BT$ from Eq.~\eqref{vdW}. Consequently, the second and the third terms in Eq.~\eqref{chem.pot.vdW} are positive and bounded.
Indeed, at $b\rho\approx 1$, the particles have very small space to move, thus small average kinetic energy and small temperature. Nevertheless, the classical limit is achieved at large temperature, otherwise, e.g., there are substantial deviations from the equipartition theorem \cite{Bialas2019,Luczka2020}.
The last term in Eq.~\eqref{chem.pot.vdW} is negative (positive) for attractive (repulsive) interactions $a>0$ ($a<0$), and its magniture is fixed by the microscopic details, $2|a|\rho\propto|\phi|b\rho\leqslant|\phi|$, and does not dominate $k_B T\ln(\rho\,\lambda_T^3)$ in the classical limit $\lambda_T^3\rho\ll1$.
In conclusion, the chemical potential of the van der Waals gas is negative.

A quantum extension of the van der Waals model \cite{Vovchenko2015,Redlich2016} is described by the equation of state

\begin{equation} \label{vdW.P}
P(T,\mu)=P_{\textnormal{id}}(T,\mu')-a\rho^2\ ,
\end{equation}
where the subscript ``$\textnormal{id}$'' denotes the functional forms of ideal (either bosonic or fermionic) gases, and
\begin{equation}
\mu'(\rho,T)=\mu_{\textnormal{id}}\left(\frac{\rho}{1-b\rho},T\right)\ .
\end{equation}
Moreover, $\mu_{\textnormal{id}}$ and $\rho_{\textnormal{id}}$ fulfil

\begin{align}
\label{vdW.mu}
& \mu'=\mu-bP_{\textnormal{id}}(T,\mu')+2a\rho\ , \\
\label{vdW.rho}
& \rho(T,\mu)=\frac{\rho_{\textnormal{id}}(T,\mu')}{1+b\,\rho_{\textnormal{id}}(T,\mu')}\ .
\end{align}
Being $\mu'$ the chemical potential of an ideal gas, it ranges over $(-\infty,0]$. Therefore, the chemical potential $\mu$ can be positive for repulsive interactions ($a<0$) or for small attractive interactions ($a>0$) if $bP_{\textnormal{id}}(T,\mu')>2a\rho\propto\phi b\rho$.

Note from Eq.~\eqref{vdW.rho} that $\rho(T,\mu)$ has a finite maximum when also $\rho_{\textnormal{id}}(T,\mu')$ is maximum. Consequently the mechanism of BEC formation is inherited from that of the ideal bosonic gas, shown in section \ref{app.ideal}, when $\mu'\to0$. The critical density of the bosonic van der Waals gas is obtained by replacing $\rho_{\textnormal{id}}$ with its maximum in Eq.~\eqref{vdW.rho}:

\begin{equation}
\rho_c=\frac{\zeta(\frac{3}{2})}{\lambda_T^2+b\,\zeta(\frac{3}{2})}
\end{equation}
that is smaller than the critical density of the ideal gas.

\section{Thermodynamic transformations} \label{app.transf}

In this section, general formulas for energy exchanges are derived.
Start with a general thermodynamic transformation in the Schr\"odinger picture, where the density matrix evolves quasistatically and is described by a grandcanonical ensemble, $\varrho=e^{-\beta(H-\mu N)}/Z$, at every time.
Plugging the grandcanonical state into $S=-k_B\textnormal{Tr}(\varrho\ln\varrho)$, one obtains the following form for the heat

\begin{equation} \label{Q}
Q=\int T\textnormal{d}S=\int\textnormal{Tr}\big((H-\mu N)\,\textnormal{d}\varrho\big)\ .
\end{equation}
In isothermal transformations, as those considered here the heat exchange is

\begin{align}
Q= & T(S_{\textnormal{f}}-S_{\textnormal{i}}) \nonumber \\
= & U_{\textnormal{f}}-U_{\textnormal{i}}+P_{\textnormal{f}}V_{\textnormal{f}}-P_{\textnormal{i}}V_{\textnormal{i}}-\mu_{\textnormal{f}}\mathcal{N}_{\textnormal{f}}+\mu_{\textnormal{i}}\mathcal{N}_{\textnormal{i}}\ ,
\end{align}
where the subscripts ``$\textnormal{i}$'' and ``$\textnormal{f}$'' indicate respectively the initial and the final quantities of the transformation.
The Hamiltonian $H$ and the number operator $N$ depend on the volume, through their eigenvectors and the eigenvalues of $H$, but do not depend on the temperature and on the chemical potential which are the thermodynamic forces. Consequently, the chemical work is

\begin{align} \label{WC}
W^{\textnormal{C}}= & -\int\mu\,\textnormal{d}\mathcal{N}=-\int\mu\textnormal{Tr}(N\textnormal{d}\varrho+\varrho\,\textnormal{d}N) \nonumber \\
= & -\int\mu\left(\textnormal{Tr}(N\textnormal{d}\varrho)+\textnormal{Tr}\left(\varrho\,\frac{\partial N}{\partial V}\right)\textnormal{d}V\right)\ .
\end{align}
Using $U=\textnormal{Tr}(H\varrho)$, the first law of thermodynamics $\Delta U=Q-W^{\textnormal{M}}-W^{\textnormal{C}}$, expressions \eqref{Q} and \eqref{WC},
one derives the following expression for the mechanical work

\begin{align} \label{WM}
W^{\textnormal{M}}= & -\int\textnormal{Tr}\big(\varrho\,(\textnormal{d}H-\mu\,\textnormal{d}N)\big) \nonumber \\
= & -\int\textnormal{Tr}\left(\varrho\,\left(\frac{\partial H}{\partial V}-\mu\,\frac{\partial N}{\partial V}\right)\right)\textnormal{d}V\ .
\end{align}

Apply now the above equations to work exchanges for the reversible transformations exploited in the cycles under considerations.

\paragraph{Transformation at constant temperature $T$ and constant chemical potential $\mu=\mu_{\textnormal{i}}=\mu_{\textnormal{f}}$:}

\begin{align}
W^{\textnormal{C}}_{\textnormal{iso-}T\mu}= & -\mu\,(\mathcal{N}_{\textnormal{f}}-\mathcal{N}_{\textnormal{i}})\ , \\
W^{\textnormal{M}}_{\textnormal{iso-}T\mu}= & P_{\textnormal{f}}V_{\textnormal{f}}-P_{\textnormal{i}}V_{\textnormal{i}}\ .
\end{align}

\paragraph{Transformation at constant temperature $T$ and constant particle number $\mathcal{N}=\mathcal{N}_{\textnormal{i}}=\mathcal{N}_{\textnormal{f}}$:}

\begin{align}
W^{\textnormal{C}}_{\textnormal{iso-}T\mathcal{N}}= & 0\ , \\
W^{\textnormal{M}}_{\textnormal{iso-}T\mathcal{N}}= & P_{\textnormal{f}}V_{\textnormal{f}}-P_{\textnormal{i}}V_{\textnormal{i}}-(\mu_{\textnormal{f}}-\mu_{\textnormal{i}})\mathcal{N}\ .
\end{align}

\paragraph{Transformation at constant temperature $T$ and constant volume $V=V_{\textnormal{i}}=V_{\textnormal{f}}$:}

\begin{align}
W^{\textnormal{C}}_{\textnormal{iso-}TV}= & (P_{\textnormal{f}}-P_{\textnormal{i}})V-\mu_{\textnormal{f}}\,\mathcal{N}_{\textnormal{f}}+\mu_{\textnormal{i}}\,\mathcal{N}_{\textnormal{i}}\ , \\
W^{\textnormal{M}}_{\textnormal{iso-}TV}= & 0\ .
\end{align}

Equations \eqref{WC} and \eqref{WM} and the identification of heat with the right-hand-side of equation \eqref{Q} are extended to non-equilibrium transformations. These energy exchanges generalise the usual definitions of quantum thermodynamics \cite{Alicki1979} to the presence of particle fluxes, and are analogous to those discussed in \cite{Cuetara2016}. There, however,
the number operator $N$ does not depend
on the external driving, whereas the engines considered in this paper require to account
for the changes of $N$ with the volume.

\section{Irreversible transformations} \label{app.irr}

In this section, irreversible transformations are modelled by finite-time dynamics of the engine substance interacting with a grandcanonical thermal bath. In the limit of infinite time  these transformations reduce to reversible, quasistatic transformations. The total Hamiltonian is $H+H_B+H_I$, where $H=\sum_{\bm p}\varepsilon_p a_{\bm p}^\dag a_{\bm p}$ is the substance Hamiltonian for ideal homogeneous gases with $\varepsilon_p=(p_x^2+p_y^2+p_z^2)/(2m)=p^2/(2m)$, momenta ${\bm p}=(p_x,p_y,p_z)=2\pi\hslash\,(n_x/L_x,n_y/L_y,n_z/L_z)$ with $n_{x,y,z}\in\mathbbm{Z}$, and canonical commutation relations $[a_{\bm p},a_{{\bm p}'}^\dag]=\delta_{{\bm p},{\bm p}'}$.

The bath is made of harmonic oscillators with Hamiltonian $H_B=\sum_{{\bm m}\in\mathbbm{N}^3}\hslash\,{\bm m}\cdot{\bm \Omega}\,b_{\bm m}^\dag b_{\bm m}$, with ${\bm m}=(m_x,m_y,m_z)$, ${\bm \Omega}=(\Omega_x,\Omega_y,\Omega_z)$, and discrete modes ($[b_{\bm m},b_{\bm m}'^\dag]=\delta_{{\bm m},{\bm m}'}$). In the continuum limit $\Omega_{x,y,z}\to0$
with the new variables $\omega_{x,y,z}=m_{x,y,z}\Omega_{x,y,z}$, ${\bm \omega}=(\omega_x,\omega_y,\omega_z)$,
one defines the continuous bosonic modes $b_{\bm \omega}=b_{\bm m}/\sqrt{\Omega_x\Omega_y\Omega_z}$ ($[b_{\bm\omega},b_{{\bm\omega}'}^\dag]\to\delta({\bm\omega}-{\bm\omega}')$),
and the bath Hamiltonian becomes $H_B\to\hslash\int\textnormal{d}{\bm\omega} \, (\omega_x+\omega_y+\omega_z) \, b_{\bm\omega}^\dag b_{\bm\omega}$.

The substance-bath interaction is bilinear in the system and bath bosonic operators:
$H_I=\lambda\sum_{\bm p}\int\textnormal{d}{\bm \omega}\,h(\hslash\omega_x+\hslash\omega_y+\hslash\omega_z)\,T_{{\bm\omega},{\bm p}}\big(a_{\bm p}^\dag b_{\bm\omega}+b_{\bm\omega}^\dag a_{\bm p}\big)$.
In order to simply the computations in the following, consider that there is an injective function ${\bm w}:{\bm p}\to
{\bm w}({\bm p})$ with $w_x({\bm p})+w_y({\bm p})+w_z({\bm p})=\varepsilon_p/\hslash$.
When $\bm \omega$ is in the image of ${\bm w}$, i.e. there exists a vector $\bm p$ with $\bm \omega={\bm w}(\bm p)$, then

\begin{align}
& \sum_{\substack{{\bm p}\\\varepsilon_p=\varepsilon}}T_{{\bm w(\bm p')},{\bm p}}\overline{T_{{\bm w(\bm p'')},{\bm p}}}=0 \quad \textnormal{if} \quad \bm p'\neq\bm p''\ , \\
& M_{\bm p'}=\sum_{\substack{{\bm p}\\\varepsilon_p=\varepsilon}}\left|T_{{\bm w(\bm p')},{\bm p}}\right|^2\sim\sum_{\substack{{\bm p}\\\varepsilon_p=\varepsilon}}1
\equiv\mathcal{M}_\varepsilon\ , \label{Mepsilon}
\end{align}
where the bar stands for complex conjugation.
Moreover, $T_{{\bm\omega},{\bm p}}$ are arbitrary if $\omega_x+\omega_y+\omega_z\neq\varepsilon_p/\hslash$.
When $\omega_x+\omega_y+\omega_z=\varepsilon_p/\hslash$ but $\bm\omega$ is not in the image of ${\bm w}$, then $T_{{\bm\omega},{\bm p}}=0$: this set of $\bm \omega$ values has vanishing measure when the moments $\bm p$ vary continuously (infinitely large size $L_{x,y,z}$), because the function $\bm w$ becomes surjective in $\mathbbm{R}^3$.
A concrete example
is provided by coefficients $T_{{\bm\omega},{\bm p}}$ borrowed from permutationally invariant generalizations of Jacobi coordinates in many-body problems \cite{Amirim2019}.

\subsection{Master equation} \label{app.master}

Consider the bath in the grandcanonical state and so large that it is not substantially perturbed by the interaction with the substance. The master equation of the substance dynamics is then derived by tracing out the bath degrees of freedom and applying the standard weak coupling regime (Born, Markov, and secular approximations
\cite{BreuerPetruccione,BenattiFloreanini}:

\begin{widetext}
\begin{align} \label{master.eq}
\frac{\textnormal{d}\varrho}{\textnormal{d}t}=\mathbb{L}[\varrho]=-\frac{i}{\hslash}\big[H+\lambda^2 H_{LS},\varrho\big]
&
+\sum_{\substack{{\bm p},{\bm p}',{\bm p}''\\\varepsilon_p=\varepsilon_{p'}\\=\varepsilon_{p''}}}\frac{\lambda^2}{\hslash^2}\,\gamma_p\big(n(\varepsilon_p)+1\big)
T_{{\bm w}({\bm p}''),{\bm p}}
T_{{\bm w}({\bm p}''),{\bm p}'}^\dag
\bigg(a_{\bm p}\varrho \, a_{{\bm p}'}^\dag-\frac{1}{2}\left\{a_{{\bm p}'}^\dag a_{\bm p},\varrho\right\}\bigg)
\nonumber \\
&
+\sum_{\substack{{\bm p},{\bm p}',{\bm p}''\\\varepsilon_p=\varepsilon_{p'}\\=\varepsilon_{p''}}}\frac{\lambda^2}{\hslash^2}\,\gamma_p\,n(\varepsilon_p)
T_{{\bm w}({\bm p}''),{\bm p}}^\dag
T_{{\bm w}({\bm p}''),{\bm p}'}
\bigg(a_{\bm p}^\dag\varrho \, a_{{\bm p}'}-\frac{1}{2}\left\{a_{{\bm p}'}a_{\bm p}^\dag,\varrho\right\}\bigg)\ ,
\end{align}
\end{widetext}
where $\gamma_p=2\pi h^2(\varepsilon_p)$, the occupation number of the bath mode with energy $\varepsilon$ is $n(\varepsilon)=\big(e^{\beta(\varepsilon-\mu)}-1\big)^{-1}$, and

\begin{equation}
H_{LS}=\sum_{\substack{{\bm p},{\bm p}',{\bm p}''\\\varepsilon_p=\varepsilon_{p'}\\=\varepsilon_{p''}}}
\Delta_p\,
T_{{\bm w}({\bm p}''),{\bm p}}
T_{{\bm w}({\bm p}''),{\bm p}'}^\dag
\,a_{\bm p}^\dag a_{{\bm p}'}
\end{equation}
is the Lamb shift Hamiltonian induced by the interaction with the bath with ($\fint$ stands for the Cauchy principal value)

\begin{equation}
\Delta_p=\fint\textnormal{d}\omega\,\frac{h^2(\hslash\omega)}{\varepsilon_p-\hslash\omega}\ .
\end{equation}

In order to decouple the dynamical degrees of freedom, it is convenient to re-write equation \eqref{master.eq} in terms of the rotated modes defined by the bosonic operators

\begin{equation} \label{A.mode}
A_{\bm p}=\frac{1}{\displaystyle
\sqrt{M_{\bm p}}
}\,
\sum_{\substack{{\bm p}'\\\varepsilon_{p'}=\varepsilon_p}}
T_{{\bm w}({\bm p}),{\bm p}'}
\,a_{{\bm p}'}
\end{equation}
with $M_{\bm p}=\mathcal{O}\big(\mathcal{M}_{\varepsilon_p}\big)$ from equation \eqref{Mepsilon},
and $\big[A_{\bm p},A_{{\bm p}'}^\dag\big]=\delta_{{\bm p},{\bm p}'}$.
One then computes

\begin{align}
\label{master.eq2}
& \frac{\textnormal{d}\varrho}{\textnormal{d}t}=\mathbb{L}[\varrho]=-\frac{i}{\hslash}\big[H+\lambda^2 H_{LS},\varrho\big]+\frac{\lambda^2}{\hslash^2}\sum_{\bm p}\gamma_p \,M_{\bm p}\times \nonumber \\
& \times \Bigg(\big(n(\varepsilon_p)+1\big)
\bigg(A_{\bm p}\varrho \, A_{\bm p}^\dag-\frac{1}{2}\left\{A_{\bm p}^\dag A_{\bm p},\varrho\right\}\bigg) \nonumber \\
& +n(\varepsilon_p)
\bigg(A_{\bm p}^\dag\varrho \, A_{\bm p}-\frac{1}{2}\left\{A_{\bm p} A_{\bm p}^\dag,\varrho\right\}\bigg)\Bigg)\ , \\
& H_{LS}=\sum_{\bm p} \Delta_p\,M_{\bm p}\,A^\dag_{\bm p}A_{\bm p}\ , \\
& H=\sum_{\bm p} \varepsilon_p\,A_{\bm p}^\dag A_{\bm p}\ ,
\end{align}
which describe
the dissipative dynamics of independent bosonic modes
whose steady state is the grandcanonical state \cite{Schaller2011}.

In thermochemical engines the system volume $V(t)$ (and $L_x(t)$, $L_y(t)$, $L_z(t)$), the chemical potential $\mu(t)$, and the inverse temperature $\beta(t)$ of the bath change in time, so that the master equation generator $\mathbb{L}$ is time-dependent (similar to those derived in \cite{Albash2012,Albash2015,Yamaguchi2017,Dann2018,Amato2019})
 and the steady state at time $t$ is the time-dependent grandcanonical state $\varrho_{\textnormal{rev}}(t)=e^{-\beta(t)(H(t)-\mu(t)N(t))}/\textnormal{Tr}\,e^{-\beta(t)(H(t)-\mu(t)N(t))}$.
Remind that the Hamiltonian $H$ and the number operator $N$ depend on the volume, through their eigenvectors and the eigenvalues of $H$, and therefore vary in time.
Slow dynamics described by the master equation \eqref{master.eq} approximates equilibrium quasistatic transformations if $\varrho_{\textnormal{rev}}(t)$ is the unique instantaneous steady state.
Note however that if $\gamma_{\tilde p}=0$ for some value $\tilde p$, also the instantaneous eigenstates of the Hamiltonian $H$ where all particles occupy modes with momenta $\bm p$ and $p^2=\tilde p^2$ are istantaneous steady states. We therefore assume $\gamma_p\neq0$ for all finite $p$, in order to ensure that $\varrho_{\textnormal{rev}}(t)$ is the unique instantaneous steady state at time $t$.

Before deriving the dynamics at long times, the next subsection introduces some approximations for large volumes that will be used later, e.g., for approximating the volume derivatives of $H$ and $N$.

\subsection{Continuum approximation} \label{app.cont.approx}

The limit of large size allows for approximating the discrete momenta ${\bm p}=2\pi\hslash\,(n_x/L_x,n_y/L_y,n_z/L_z)$ are with continuous variable ${\bm p}=(p_x,p_y,p_z)\in\mathbbm{R}^3$.
 one then estimates the unconstrained sum of ${\bm p}$ with arbitrary operators $f$:

\begin{equation}  \label{cont.sum0}
\sum_{\bm p}f({\bm p})\simeq\frac{V}{(2\pi\hslash)^3}\int\textnormal{d}{\bm p}\,f({\bm p})
\ .
\end{equation}

Applications are

\begin{align}
H\simeq & \,\frac{V}{(2\pi\hslash)^3}\int_{\mathbbm{R}^3}\textnormal{d}{\bm p}\,\varepsilon_p\,a_{\bm p}^\dag a_{\bm p}\ , \\
N\simeq & \,\frac{V}{(2\pi\hslash)^3}\int_{\mathbbm{R}^3}\textnormal{d}{\bm p}\,a_{\bm p}^\dag a_{\bm p}\ .
\end{align}
Within the continuum approximation, momenta are continuous variables that range in a volume independent domains, i.e., $\mathbbm{R}^3$, so that the energy $\varepsilon_p$ as well as the creation and annihilation operators, $a_{\bm p}^\dag$ and $a_{\bm p}$, no longer depend on the volume. Therefore, this approximation
provides a simple estimation of following volume derivatives,

\begin{align}
\label{Hder}
\frac{\partial H}{\partial V} & \simeq\frac{H}{V}\ , \\
\label{Nder}
\frac{\partial N}{\partial V} & \simeq\frac{N}{V}\ .
\end{align}

The sum of ${\bm p}'$ constrained to
$2m\varepsilon_{p'}=(2\pi\hslash)^2\big(n_x'^2/L_x^2+n_y'^2/L_y^2+n_z'^2/L_z^2\big)=p^2=2m\varepsilon_p$, as in the master equation \eqref{master.eq},
is approximated with the surface integral over the ellipsoid with semi-axes $(2\pi\hslash)^2/(pL_x)^2$, $(2\pi\hslash)^2/(pL_y)^2$, and $(2\pi\hslash)^2/(pL_z)^2$, when the energy $\varepsilon_p$ is larger than $\lambda_T^2/(\beta L_z^2)$. This energy value is the energy contribution due to the smallest non-zero component of the momentum along the $z$ axis. Below this energy, particles have vanishing momenta along the $z$ axis, and the sum of ${\bm p}'$ is constrained to
$(2\pi\hslash)^2\big(n_x'^2/L_x^2+n_y'^2/L_y^2/L_z^2\big)=p^2$, and approximated with the surface integral over the ellipse with semi-axes $(2\pi\hslash)^2/(pL_x)^2$ and $(2\pi\hslash)^2/(pL_y)^2$ when $\varepsilon_p\geqslant\lambda_T^2/(\beta L_y^2)$. If $\varepsilon_p<\lambda_T^2/(\beta L_y^2)$, also momenta along the $y$ axis vanish, and the sum of ${\bm p}'$ is constrained to
$\big(2\pi\hslash n_x'/L_x\big)^2=p^2$, when $\varepsilon_p\geqslant\lambda_T^2/(\beta L_x^2)$. All particles with energy smaller than $\lambda_T^2/(\beta L_x^2)$ have vanishing momenta
such that $\varepsilon_p=\varepsilon_{p'}=0$. Therefore, one obtains

\begin{widetext}
\begin{align} \label{cont.sum}
& \sum_{\substack{{\bm p}'\\\varepsilon_p=\varepsilon_{p'}}}f({\bm p},{\bm p}')
=\sum_{n'_x,n'_y,n'_z}f({\bm p},{\bm p}')\,\delta_{(2\pi\hslash)^2\left(\frac{n'^2_x}{L_x^2}+\frac{n'^2_y}{L_y^2}+\frac{n'^2_z}{L_z^2}\right),\,p^2} \nonumber \\
& =f({\bm 0},{\bm 0})
+\sum_{n'_x\neq0}f({\bm p},{\bm p}')\,\delta_{\left(\frac{2\pi\hslash n'_x}{L_x}\right)^2,\,p^2}+\sum_{n'_x,n'_y\neq0}f({\bm p},{\bm p}')\,\delta_{(2\pi\hslash)^2\left(\frac{n'^2_x}{L_x^2}+\frac{n'^2_y}{L_y^2}\right),\,p^2}+\sum_{n'_x,n'_y,n'_z\neq0}f({\bm p},{\bm p}')\,\delta_{(2\pi\hslash)^2\left(\frac{n'^2_x}{L_x^2}+\frac{n'^2_y}{L_y^2}+\frac{n'^2_z}{L_z^2}\right),\,p^2} \nonumber \\
& \simeq
\begin{cases}
\displaystyle
\frac{L_xL_y\,p^2}{(2\pi\hslash)^2}\int_0^\pi\sin\vartheta\,\textnormal{d}\vartheta \int_0^{2\pi}\textnormal{d}\varphi\,f({\bm p},{\bm p}')
\sqrt{\cos^2\vartheta+\frac{L_z^2}{L_x^2}\,\sin^2\vartheta\cos^2\varphi+\frac{L_z^2}{L_y^2}\,\sin^2\vartheta\sin^2\varphi}
& \displaystyle \textnormal{if} \quad \varepsilon_p\geqslant\frac{\lambda_T^2}{\beta L_z^2} \\
\displaystyle \frac{L_x\,p}{2\pi\hslash}\int_0^{2\pi}\textnormal{d}\varphi\,f({\bm p},{\bm p}')\,\sqrt{\sin^2\varphi+\frac{L_y^2}{L_x^2}\,\cos^2\varphi}
& \displaystyle \textnormal{if} \quad \frac{\lambda_T^2}{\beta L_y^2}\leqslant\varepsilon_p<\frac{\lambda_T^2}{\beta L_z^2} \\
\displaystyle f({\bm p},{\bm p})+f({\bm p},-{\bm p})
& \displaystyle \textnormal{if} \quad \frac{\lambda_T^2}{\beta L_x^2}\leqslant\varepsilon_p<\frac{\lambda_T^2}{\beta L_y^2} \\
\displaystyle
f({\bm 0},{\bm 0}) & \textnormal{if} \quad \varepsilon_p=0
\end{cases}\ ,
\end{align}
\end{widetext}
where ${\bm p}'$ is identified in spherical coordinates by the polar and the azimuthal angles with respect to the equatorial $x$-$y$ plane, $\vartheta$ and $\varphi$ respectively, and the modulus $p'=p$.
Remind that $L_x\geqslant L_y\geqslant L_z$, such that the square roots in equation \eqref{cont.sum} are bounded. The careful treatment of the energy scales $\lambda_T^2/(\beta L_{x,y,z}^2)$ allows us to identify the relevant contributions to heat and work corrections in irreversible transformations with $d$D-BECs.
Equation \eqref{cont.sum} is also used to approximate $\mathcal{M}_{\varepsilon_p}$ and thus to estimate the size scaling of $M_{\bf p}$ in equation \eqref{Mepsilon}. Indeed using $f({\bm p},{\bm p}')=1$ in equation \eqref{cont.sum}, one obtains

\begin{equation} \label{cont.sum.gamma}
\mathcal{M}_{\varepsilon_p}
\simeq
\begin{cases}
\displaystyle
\frac{L_xL_y\,p^2}{(2\pi\hslash)^2}\,C
& \displaystyle \textnormal{if} \quad \varepsilon_p\geqslant\frac{\lambda_T^2}{\beta L_z^2} \\
\displaystyle \frac{L_x\,p}{2\pi\hslash}\,G
& \displaystyle \textnormal{if} \quad \frac{\lambda_T^2}{\beta L_y^2}\leqslant\varepsilon_p<\frac{\lambda_T^2}{\beta L_z^2} \\
\displaystyle 2
& \displaystyle \textnormal{if} \quad \frac{\lambda_T^2}{\beta L_x^2}\leqslant\varepsilon_p<\frac{\lambda_T^2}{\beta L_y^2} \\
\displaystyle
1 & \textnormal{if} \quad \varepsilon_p=0
\end{cases}\ ,
\end{equation}
with

\begin{align}
C= & \int_0^\pi\sin\vartheta\,\textnormal{d}\vartheta \int_0^{2\pi}\textnormal{d}\varphi \nonumber \\
& \sqrt{\cos^2\vartheta+\frac{L_z^2}{L_x^2}\,\sin^2\vartheta\cos^2\varphi+\frac{L_z^2}{L_y^2}\,\sin^2\vartheta\sin^2\varphi}\ , \\
G= & \int_0^{2\pi}\textnormal{d}\varphi\,\sqrt{\sin^2\varphi+\frac{L_y^2}{L_x^2}\,\cos^2\varphi}\ .
\end{align}

\subsection{Instantaneous relaxation times} \label{app.decay}

The instantaneous relaxation time is the time the dynamics takes to approach the steady state as if the master equation \eqref{master.eq} have time-independent coefficients evaluated at $t$.
From the structure of
single-mode
master equations that are quadratic in the bosonic operators \cite{Sandulescu1987,Isar1994}, as in equation \eqref{master.eq2}, the relaxation time of each mode $(A_{\bm p},A_{\bm p}^\dag)$ is
$\theta_{\bm p}(t)=\hslash^2/\big(\lambda^2\,\gamma_p\,M_{\bm p}\big)$.
Remind that $M_{\bm p}\sim\mathcal{M}_{\varepsilon_p}$ from equation \eqref{Mepsilon},
and the size scaling of $M_{\bm p}$ is given by equation \eqref{cont.sum.gamma}. Therefore, the instantaneous relaxation times for $\varepsilon_p\geqslant\frac{\lambda_T^2}{\beta L_z^2}$ are estimated at large size by

\begin{align} \label{theta}
\theta_{\bm p}\sim &
\frac{4\pi^2\hslash^4}{\lambda^2\,\gamma_p\,p^2\,L_xL_y}\left(\int_0^\pi\sin\vartheta\,\textnormal{d}\vartheta\int_0^{2\pi}\textnormal{d}\varphi\bigg(\cos^2\vartheta \right. \nonumber \\
& \left.
+\frac{L_z^2}{L_x^2}\,\sin^2\vartheta\cos^2\varphi+\frac{L_z^2}{L_y^2}\,\sin^2\vartheta\sin^2\varphi\bigg)^{\frac{1}{2}}\right)^{-1}\ .
\end{align}
For later convenience, define also the maximum relaxation time $\bar\theta=\max_{{\bm p},t}\theta_{\bm p}(t)$ that scales as $\mathcal{O}(L_z/V)$.

The number of bosonic modes with energy $\varepsilon_p<\frac{\lambda_T^2}{\beta L_z^2}$ is vanishingly small for large size and provide negligible contributions to the substance state and to instantaneous relaxation times \eqref{theta} without BECs.
These contributions are negligible also in 2D-BEC or 1D-BEC phases where the condensates consist of all momenta lying respectively in the $x$-$y$ plane or along the $x$ direction. The 0D-BEC consists only of the ground state that contributes to thermodynamic quantities with vanishing energy $\varepsilon_0=0$, and with particle number relaxing to $\langle a_{\bm 0}^\dag a_{\bm 0}\rangle=f\mathcal{N}=\mathcal{N}-V\rho_c$ when the number of non-condensed particles approaches the critical value $V\rho_c$.
As equation \eqref{theta} gives the relevant instantaneous relaxation times for almost all excited states (except the aforementioned vanishingly small number of modes), it does also for 0D-BECs.

\subsection{Time-evolution} \label{app.evol}

For infinitely slow dynamics compared to the instantaneous relaxation rates, relaxation to $\rho_{\textnormal{rev}}(t)$ at each $t$ happens in negligible time, recovering an equilibrium quasistatic transformation.
Given a large but finite total time of the dynamics $\tau$, the time-evolution is a perturbation of the quasistatic transformation \cite{Cavina2017}.
Consider then the following perturbative expansion

\begin{equation} \label{expans.slow}
\varrho(t)=\varrho_{\textnormal{rev}}(t)+\frac{\bar\theta}{\tau}\,\varrho_{\textnormal{irr}}(t)+\mathcal{O}\left(\frac{\bar\theta}{\tau}\right)^2\ .
\end{equation}
Defining $t'=t/\tau\in[0,1]$, the master equation reads $\textnormal{d}\varrho/\textnormal{d}t'=\tau\,\mathbb{L}\big[\varrho\big]$.
Plugging the expantion \eqref{expans.slow} into this master equation provides a self-consistency condition:

\begin{equation} \label{self}
\frac{\textnormal{d}\varrho_{\textnormal{rev}}}{\textnormal{d}t'}=\bar\theta\,\mathbb{L}\big[\varrho_{\textnormal{irr}}\big]\ .
\end{equation}

The left-hand-side of the self-consistency condition \eqref{self} is

\begin{align} \label{self.left}
& \frac{\textnormal{d}\varrho_{\textnormal{rev}}}{\textnormal{d}t'}=\frac{\partial\mu}{\partial t'}\,\beta\Big(N-\textnormal{Tr}\big(N\varrho_{\textnormal{rev}}\big)\Big)\varrho_{\textnormal{rev}} \nonumber \\
& -\frac{\partial V}{\partial t'}\,\beta\Bigg(\frac{\partial(H-\mu N)}{\partial V}-\textnormal{Tr}\bigg(\varrho_{\textnormal{rev}}\frac{\partial(H-\mu N)}{\partial V}\bigg)\Bigg)\varrho_{\textnormal{rev}} \nonumber \\
& -\frac{\partial\beta}{\partial t'}\Big(H-\mu N-\textnormal{Tr}\big(\big(H-\mu N\big)\varrho_{\textnormal{rev}}\big)\Big)\varrho_{\textnormal{rev}}\ .
\end{align}
The contribution proportional to $\partial_{t'}\beta$ vanishes in isothermal transformations, as those considered in this paper.
The volume derivatives within the continuum approximation are expressed in equations \eqref{Hder} and \eqref{Nder}.

The self-consistency condition \eqref{self} will be exploited to derive the first order correction to the quasistatic evolution, namely $\varrho_{\textnormal{irr}}$.
First, write the general form of $\varrho_{\textnormal{irr}}$, by noting that the master equation \eqref{master.eq} is quadratic in the field operators $\{a_{\bm p},a_{\bm p}^\dag\}_{\bm p}$, as well as in $\{A_{\bm p},A_{\bm p}^\dag\}_{\bm p}$, and therefore preserves the gaussianity of $\varrho$.
Consequently, the exact state can be written as

\begin{equation} \label{gauss}
\varrho(t)=\frac{\displaystyle \exp\left(-\beta H(t)+\beta\mu(t) N(t)+\frac{\beta\,\bar\theta}{\tau}\, X(t)\right)}{\displaystyle \textnormal{Tr}\,\exp\left(-\beta H(t)+\beta\mu(t) N(t)+\frac{\beta\,\bar\theta}{\tau}\, X(t)\right)}\ ,
\end{equation}
with a quadratic Hermitian operator $X=\sum_{{\bm p},{\bm p}'}\chi_{{\bm p},{\bm p}'}A_{\bm p}^\dag A_{{\bm p}'}$.
Expanding \eqref{gauss} in powers of $\bar\theta/\tau$ \cite{Wilcox1967},

\begin{align}
& e^{-\beta\left(H-\mu N-\frac{\bar\theta}{\tau}X\right)}=e^{-\beta(H-\mu N)} \nonumber \\
& \qquad +\frac{\bar\theta}{\tau}\int_0^\beta\textnormal{d}s\,e^{-s(H-\mu N)}Xe^{-(\beta-s)(H-\mu N)} \nonumber \\
& \qquad +\mathcal{O}\left(\frac{\bar\theta}{\tau}\right)^2\ ,
\nonumber \\
& \textnormal{Tr}\,e^{-\beta\left(H-\mu N-\frac{\bar\theta}{\tau}X\right)}=\textnormal{Tr}\left(e^{-\beta(H-\mu N)}\right) \nonumber \\
& \qquad +\frac{\beta\,\bar\theta}{\tau}\,\textnormal{Tr}\left(Xe^{-\beta(H-\mu N)}\right)+\mathcal{O}\left(\frac{\bar\theta}{\tau}\right)^2\ ,
\nonumber
\end{align}
and using canonical commutation relations of bosonic operators, $\big[A_{\bm p},A_{{\bm p}'}^\dag\big]=\delta_{{\bm p},{\bm p}'}$,
one obtains for $\varrho_{\textnormal{irr}}$

\begin{align} \label{rho.irr}
\varrho_{\textnormal{irr}}\simeq\ & \tau\,(\varrho-\varrho_{\textnormal{rev}})\simeq-\beta\,\textnormal{Tr}\left(X\varrho_{\textnormal{rev}}\right)\varrho_{\textnormal{rev}} \nonumber \\
& + \frac{1}{Z_{\textnormal{rev}}}\int_0^\beta\textnormal{d}s\,e^{-s(H-\mu N)}X\,e^{-(\beta-s)(H-\mu N)} \nonumber \\
= & \sum_{{\bm p},{\bm p}'}\frac{e^{\beta(\varepsilon_{p'}-\varepsilon_p)}-1}{\varepsilon_{p'}-\varepsilon_p}\,\chi_{{\bm p},{\bm p}'}\,A_{\bm p}^\dag A_{{\bm p}'}\,\varrho_{\textnormal{rev}} \nonumber \\
& -\beta\,\textnormal{Tr}\left(X\varrho_{\textnormal{rev}}\right)\varrho_{\textnormal{rev}}\ .
\end{align}
Using the expression \eqref{rho.irr} and after algebraic manipulations of bosonic operators, $\mathbb{L}\big[\varrho_{\textnormal{irr}}\big]$ is

\begin{eqnarray} \label{self.right}
&& -\frac{i}{\hslash}\sum_{{\bm p},{\bm k}}\frac{e^{\beta(\varepsilon_k-\varepsilon_p)}-1}{\varepsilon_k-\varepsilon_p}\,\chi_{{\bm p},{\bm k}}\,(\varepsilon_p+\lambda^2 M_{\bm p}\,\Delta_p)\,A_{\bm p}^\dag A_{\bm k}\,\varrho_{\textnormal{rev}} \nonumber \\
&& +\frac{i}{\hslash}\sum_{{\bm p},{\bm k}}\frac{e^{\beta(\varepsilon_p-\varepsilon_k)}-1}{\varepsilon_p-\varepsilon_k}\,\chi_{{\bm k},{\bm p}}\,(\varepsilon_p+\lambda^2 M_{\bm p}\,\Delta_p)\,A_{\bm k}^\dag A_{\bm p}\,\varrho_{\textnormal{rev}} \nonumber \\
&& +\frac{\lambda^2\beta}{\hslash^2}\sum_{{\bm p}}\gamma_p\,M_{\bm p}\,\chi_{{\bm p},{\bm p}}\,n(\varepsilon_p)\,\varrho_{\textnormal{rev}} \nonumber \\
&& -\frac{\lambda^2}{2\hslash^2}\sum_{{\bm p},{\bm k}}\frac{e^{\beta(\varepsilon_k-\varepsilon_p)}-1}{\varepsilon_k-\varepsilon_p}\,\gamma_p\,M_{\bm p}\,\chi_{{\bm p},{\bm k}}\,A_{\bm p}^\dag A_{\bm k}\,\varrho_{\textnormal{rev}} \nonumber \\
&& -\frac{\lambda^2}{2\hslash^2}\sum_{{\bm p},{\bm k}}\frac{e^{\beta(\varepsilon_p-\varepsilon_k)}-1}{\varepsilon_p-\varepsilon_k}\,\gamma_p\,M_{\bm p}\,\chi_{{\bm k},{\bm p}}\,A_{\bm k}^\dag A_{\bm p}\,\varrho_{\textnormal{rev}}\ .
\end{eqnarray}

Since the operators $A_{\bm p}^\dag A_{{\bm p}'}\varrho_{\textnormal{rev}}$ for all ${\bm p},{\bm p}'$ and $\varrho_{\textnormal{rev}}$ form an operator basis in the self-consistency condition \eqref{self}, $\chi_{{\bm p},{\bm p}'}$ are determined by equating the coefficients of these operators between the left- and the right-hand-side of \eqref{self}.
At this point, note that the master equation \eqref{master.eq} is symmetric with respect to rotations of momenta
${\bm p}$ and ${\bm p}'$ where $p=p'$, so is the time-evolution $\varrho(t)$ (if the initial state has the same symmetry), and thus $\chi_{{\bm p},{\bm p}'}\equiv\chi_p$ are the same for all such ${\bm p}$ and ${\bm p}'$.
Let us focus on the derivation of $\chi_{{\bm p},{\bm p}}=\chi_p$
which appears in the coefficient of the operator $A_{\bm p}^\dag A_{\bm p}\varrho_{\textnormal{rev}}$ in \eqref{self.right}.
This coefficient must be equal to the similar coefficient in equation \eqref{self.left}, according to the self-consistency condition \eqref{self}. Notice also that the substance Hamiltonian and particle number operators can be re-written as $H=\sum_{\bm p}\varepsilon_p A_{\bm p}^\dag A_{\bm p}$ and $N=\sum_{\bm p}A_{\bm p}^\dag A_{\bm p}$ respectively, since the rotated modes $A_{\bm p}$ are linear combinations of all the original modes $a_{\bm p}$ corresponding to the same energy.

The computation sketched above results in

\begin{equation} \label{self.coeff}
\chi_p=\frac{2\hslash^2}{\lambda^2\beta\,\bar\theta M_{\bm p}}\,
\big(\xi\,\mu+\widetilde\xi\,\varepsilon_p\big)\ ,
\end{equation}
where

\begin{align}
\xi= & -\frac{\beta}{V}\,\frac{\partial V}{\partial t'}-\frac{\partial\beta}{\partial t'}-\frac{\beta}{\mu}\,\frac{\partial\mu}{\partial t'}\ , \\
\widetilde\xi= & \frac{\beta}{V}\,\frac{\partial V}{\partial t'}+\frac{\partial\beta}{\partial t'}\ ,
\end{align}
depend on the bath external parameters (volume, chemical potential, and temperature) and their time derivatives. Note that $\xi$ and $\widetilde\xi$ are intensive and remain finite also in BEC phases where $\mu\to0$.

Recalling that $M_{\bm p}\sim\mathcal{M}_{\varepsilon_p}$, $M_{\bm p}$ are estimated by equation \eqref{cont.sum.gamma} whose size scalings are used later in the computation of work and heat corrections in the non-condensed and in the $d$ED-BEC phases.
Moreover, only the terms with ${\bm p}={\bm p}'$ in the expression \eqref{rho.irr} contribute to the first corrections to work and heat given by $\rho_{\textnormal{irr}}$,
as show in the next section.

\subsection{Work and heat corrections} \label{en.corr}

The heat and work exchanges of irreversible transformations are the right-hand-sides of equations \eqref{Q}, \eqref{WC} and \eqref{WM}.
Plugging there the expansion \eqref{expans.slow}, the approximations \eqref{W.irr} and \eqref{Q.irr} are dervied with the following first order coontributions in $\bar\theta/\tau$:

\begin{widetext}
\begin{align}
\label{WM.irr}
W^{\textnormal{M}}_{\textnormal{irr}}= & -\int_0^1\textnormal{d}t'\textnormal{Tr}\left(\varrho_{\textnormal{irr}}\left(\frac{\partial H}{\partial t'}-\mu\,\frac{\partial N}{\partial t'}\right)\right)\ , \\
\label{WC.irr}
W^{\textnormal{C}}_{\textnormal{irr}}= & -\int_0^1\textnormal{d}t'\mu\,\textnormal{Tr}\left(N\,\frac{\partial\varrho_{\textnormal{irr}}}{\partial t'}+\varrho_{\textnormal{irr}}\,\frac{\partial N}{\partial t'}\right)
=\mu(0)\,\textnormal{Tr}\big(\varrho_{\textnormal{irr}}(0)N(0)\big)-\mu(\tau)\,\textnormal{Tr}\big(\varrho_{\textnormal{irr}}(\tau)N(\tau)\big)
+\int_0^1\textnormal{d}t'\,\frac{\partial\mu}{\partial t'}\,\textnormal{Tr}(N\varrho_{\textnormal{irr}})\ , \\
\label{QQ.irr}
Q_{\textnormal{irr}}= & \int_0^1\textnormal{d}t'\textnormal{Tr}\left((H-\mu N)\,\frac{\partial\varrho_{\textnormal{irr}}}{\partial t'}\right) \nonumber \\
= & \,\textnormal{Tr}\Big(\big(H(\tau)-\mu(\tau)N(\tau)\big)\varrho_{\textnormal{irr}}(\tau)\Big)
-\textnormal{Tr}\Big(\big(H(0)-\mu(0)N(0)\big)\varrho_{\textnormal{irr}}(0)\Big)
-\int_0^1\textnormal{d}t'\textnormal{Tr}\left(\varrho_{\textnormal{irr}}\,\frac{\partial(H-\mu N)}{\partial t'}\right)\Bigg)\ ,
\end{align}
\end{widetext}
where integration by parts have been used.
Note that the traces involving the time derivatives of $N$ and $H$ can be written, by virtue of equations \eqref{Hder} and \eqref{Nder}, as

\begin{align}
\label{Nirr.der.t}
\textnormal{Tr}\left(\varrho_{\textnormal{irr}}\,\frac{\partial N}{\partial t'}\right)=
&  \textnormal{Tr}\left(\varrho_{\textnormal{irr}}\,\frac{\partial N}{\partial V}\right)\frac{\partial V}{\partial t'}
\simeq 
\frac{\textnormal{Tr}(N\varrho_{\textnormal{irr}})}{V}\,\frac{\partial V}{\partial t'}\ , \\
\label{Hirr.der.t}
\textnormal{Tr}\left(\varrho_{\textnormal{irr}}\,\frac{\partial H}{\partial t'}\right) = &
\textnormal{Tr}\left(\varrho_{\textnormal{irr}}\,\frac{\partial H}{\partial V}\right)\frac{\partial V}{\partial t'}
\simeq 
\frac{\textnormal{Tr}(H\varrho_{\textnormal{irr}})}{V}\,\frac{\partial V}{\partial t'}\ .
\end{align}
Therefore, the traces $\textnormal{Tr}(N\varrho_{\textnormal{irr}})$ and $\textnormal{Tr}(H\varrho_{\textnormal{irr}})$ are the only relevant traces to be computed in equations \eqref{WM.irr}, \eqref{WC.irr} and \eqref{QQ.irr}.
Since the operators $H$ and $N$
are linear combinations of $A_{\bm p}^\dag A_{\bm p}$, only the terms with ${\bm p}={\bm p}'$ in the expression \eqref{rho.irr} for $\varrho_{\textnormal{irr}}$ contribute to the computation of these traces.
Using the expression \eqref{rho.irr}, $\textnormal{Tr}\big((A_{\bm p}^\dag A_{\bm p})^2\varrho_{\textnormal{rev}}\big)=2\,n^2(\varepsilon_p)+n(\varepsilon_p)$, and the continuum approximation, one obtains

\begin{align}
\label{tr.N.irr}
\textnormal{Tr}(N\varrho_{\textnormal{irr}}) & =\beta\,\sum_{\bm p} \chi_p\big(n^2(\varepsilon_p)+n(\varepsilon_p)\big) \nonumber \\
& \simeq\frac{\beta\,V}{(2\pi\hslash)^3}\int\textnormal{d}{\bm p}\,\chi_p\big(n^2(\varepsilon_p)+n(\varepsilon_p)\big) \nonumber \\
& +\frac{\beta\,V^{d\textnormal{D}}}{(2\pi\hslash)^d}\int\textnormal{d}{\bm p}\,
\delta({\bm p}-{\bm p}^{d\textnormal{D}})\,
\chi_p\big(n^2(\varepsilon_p)+n(\varepsilon_p)\big)\ , \\
\label{tr.H.irr}
\textnormal{Tr}(H\varrho_{\textnormal{irr}}) & =\beta\,\sum_{\bm p} \varepsilon_p\,\chi_p\big(n^2(\varepsilon_p)+n(\varepsilon_p)\big) \nonumber \\
& \simeq\frac{\beta\,V}{(2\pi\hslash)^3}\int\textnormal{d}{\bm p}\,\varepsilon_p\,\chi_p\big(n^2(\varepsilon_p)+n(\varepsilon_p)\big) \nonumber \\
& +\frac{\beta\,V^{d\textnormal{D}}}{(2\pi\hslash)^d}\int\textnormal{d}{\bm p}\,
\delta({\bm p}-{\bm p}^{d\textnormal{D}})\,
\varepsilon_p\,\chi_p\big(n^2(\varepsilon_p)+n(\varepsilon_p)\big)\ .
\end{align}
The last terms in equations \eqref{tr.N.irr} and \eqref{tr.H.irr} are the contributions of the $d$D-BEC, as described in appendix \ref{app.ideal}, formed of energy eigenstates with momenta ${\bm p}^{d\textnormal{D}}$
where ${\bm p}^{2\textnormal{D}}=(p_x,p_y,0)$ are momenta in the $x$-$y$ plane, ${\bm p}^{1\textnormal{D}}=(p_x,0,0)$ are momenta along the $x$ axis, and ${\bm p}^{0\textnormal{D}}=(0,0,0)$,
$V^{2\textnormal{D}}=L_xL_y$, $V^{1\textnormal{D}}=L_x$, and $V^{0\textnormal{D}}=1$. These contributions emerge only after the trace and not in the operators because of the singularity of bosonic occupancies when $e^{\beta\mu}\to1$. On the other hand, if $e^{\beta\mu}\neq 1$, the BEC contributions are negligible.

After plugging the coefficient $\chi_p$ \eqref{self.coeff} into equations \eqref{tr.N.irr} and \eqref{tr.H.irr}, one has to assume the functional form of $\gamma_p$ in order to compute the integrals.
Therefore, consider the general power series $\gamma_p^{-1}=\sum_j\kappa_j\,\varepsilon_p^{\alpha_j}$.
Furthermore, the change of variable $p\to\varepsilon_p=:\varepsilon$ in equations \eqref{tr.N.irr} and \eqref{tr.H.irr} results in the following types of integrals:

\begin{align}
\label{change.var3}
& \int\textnormal{d}{\bm p}\,f(p)=(2\,m)^{\frac{3}{2}}\,\pi\int_0^\infty\textnormal{d}\varepsilon\,\sqrt{\varepsilon}\,f\big(\sqrt{2m\varepsilon}\big)\ , \\
\label{change.var12}
& \int\textnormal{d}{\bm p}\,\delta({\bm p}-{\bm p}^{d\textnormal{D}})\,f(p)=\frac{(2\pi m)^{\frac{d}{2}}}{\Gamma\left(\frac{d}{2}\right)}\int_0^\infty\textnormal{d}\varepsilon\,\varepsilon^{\frac{d}{2}-1}\,f\big(\sqrt{2m\varepsilon}\big)\ , \\
\label{change.var0}
& \int\textnormal{d}{\bm p}\,\delta({\bm p}-{\bm p}^{0\textnormal{D}})\,f(p)=f(0)\ ,
\end{align}
with $d=1,2$ in equation \eqref{change.var12}.
The traces \eqref{tr.N.irr} and \eqref{tr.H.irr} are then linear combinations of the following prototypical integrals

\begin{widetext}
\begin{equation} \label{prot.int}
I_a^A=\int_A^\infty\textnormal{d}\varepsilon\,\varepsilon^a\big(n^2(\varepsilon)+n(\varepsilon)\big)
=\int_A^\infty\textnormal{d}\varepsilon\,\varepsilon^a\sum_{j=1}^\infty j\,e^{j\beta(\mu-\varepsilon)}
=\sum_{j=1}^\infty \frac{e^{j\beta\mu}}{j^a\beta^{a+1}}\Gamma(a+1,j\beta A)
\end{equation}
where $A$ is one of the energy bounds in equation \eqref{cont.sum}, and $\Gamma(s,z)$ is the incomplete Gamma function. The series representation

\begin{equation}
\Gamma(s,z)=\int_z^\infty\textnormal{d}t\,t^{s-1}e^{-t}
=
\Gamma(s)\left(1-z^s\,e^{-z}\sum_{k=0}^\infty\frac{z^k}{\Gamma(s+k+1)}\right) \quad \textnormal{if } -s\notin\mathbbm{N}
\ ,
\end{equation}
implies the following series

\begin{equation}
I_a^A=\frac{\Gamma(a+1)}{\beta^{a+1}}\left(\textnormal{Li}_a(e^{\beta\mu})-\sum_{k=0}^\infty
\frac{(\beta A)^{a+k+1}}{\Gamma(a+k+2)}\,\textnormal{Li}_{-k-1}\big(e^{\beta(\mu-A)}\big)\right)
\quad \textnormal{if } -a\notin\mathbbm{N}^+
\end{equation}
For $a=-1$, the sum in the right-hand-side of equation \eqref{prot.int} is approximated by an integral

\begin{equation} \label{exp.I1}
I_{-1}^A=\int_{1}^\infty\textnormal{d}j \, j \, e^{j\beta\mu}\,\Gamma(0,j\beta A)
=\frac{1}{\beta^2\mu^2(A-\mu)}
\left(
\mu\,e^{\beta(\mu-A)}
+(A-\mu) \big(
(1-\beta\mu)\,e^{\beta\mu}\,\Gamma(0,\beta A)
-\Gamma(0,\beta A-\beta\mu)
\big)
\right)
\ ,
\end{equation}
where $A>0$ and $\mu<0$ for the ideal Bose gas have been used.
Define also

\begin{equation}
I_a^{A,B}=\int_A^B\textnormal{d}\varepsilon\,\varepsilon^a\big(n^2(\varepsilon)+n(\varepsilon)\big)
=I_a^A-I_a^B\ .
\end{equation}

With all these manipulations, equations \eqref{tr.N.irr} and \eqref{tr.H.irr} become

\begin{align}
\label{Nirr}
\textnormal{Tr}(N\varrho_{\textnormal{irr}})\simeq & \,
\frac{2\,\hslash^2\,\sqrt{\pi\beta}\,L_z}{\lambda^2\lambda_T\,\bar\theta}
\sum_jC\,\kappa_j\left(
\xi\,\mu\,I_{\alpha_j-\frac{1}{2}}^{\frac{\lambda_T^2}{\beta L_z^2}}+\widetilde\xi\,I_{\alpha_j+\frac{1}{2}}^{\frac{\lambda_T^2}{\beta L_z^2}}
\right)
\nonumber \\
&
+
\big(\delta_{d,2}+\delta_{d,1}\big)
\left(
\frac{2\pi\,\hslash^2\,\beta^{\frac{d}{2}-1}\,V^{d\textnormal{D}}}{\Gamma\left(\frac{d}{2}\right)\lambda^2\lambda_T^{d-2}\,\bar\theta L_x L_y}
\sum_jC\,\kappa_j\left(
\xi\,\mu\,I_{\alpha_j+\frac{d}{2}-2}^{\frac{\lambda_T^2}{\beta L_z^2}}+\widetilde\xi\,I_{\alpha_j+\frac{d}{2}-1}^{\frac{\lambda_T^2}{\beta L_z^2}}
\right)
\right. \nonumber \\
& \left. +
\frac{2\sqrt{\pi}\,\hslash^2\,\beta^{\frac{d-1}{2}}\,V^{d\textnormal{D}}}{\Gamma\left(\frac{d}{2}\right)\lambda^2\lambda_T^{d-1}\,\bar\theta L_x}
\sum_jG\,\kappa_j\left(
\xi\,\mu\,I_{\alpha_j+\frac{d-3}{2}}^{\frac{\lambda_T^2}{\beta L_y^2},\frac{\lambda_T^2}{\beta L_z^2}}+\widetilde\xi\,I_{\alpha_j+\frac{d-1}{2}}^{\frac{\lambda_T^2}{\beta L_y^2},\frac{\lambda_T^2}{\beta L_z^2}}
\right)
\right) \nonumber \\
& +
\delta_{d,1}
\frac{4\,\hslash^2\,\beta^{\frac{d}{2}}\,V^{d\textnormal{D}}}{\Gamma\left(\frac{d}{2}\right)\lambda^2\lambda_T^d\,\bar\theta}
\sum_j\kappa_j\left(
\xi\,\mu\,I_{\alpha_j+\frac{d}{2}-1}^{\frac{\lambda_T^2}{\beta L_x^2},\frac{\lambda_T^2}{\beta L_y^2}}+\widetilde\xi\,I_{\alpha_j+\frac{d}{2}}^{\frac{\lambda_T^2}{\beta L_x^2},\frac{\lambda_T^2}{\beta L_y^2}}
\right)+\delta_{d,0}
\frac{2\,\hslash^2}{\lambda^2\,\gamma_{{\bm 0},{\bm 0}}\,\bar\theta}
\,\frac{\xi\,\mu\,e^{\beta\mu}}{\big(e^{\beta\mu}-1\big)^2}
\ , \\
\label{Hirr}
\textnormal{Tr}(H\varrho_{\textnormal{irr}})= & \,
\frac{2\,\hslash^2\,\sqrt{\pi\beta}\,L_z}{\lambda^2\lambda_T\,\bar\theta}
\sum_jC\,\kappa_j\left(
\xi\,\mu\,I_{\alpha_j+\frac{1}{2}}^{\frac{\lambda_T^2}{\beta L_z^2}}+\widetilde\xi\,I_{\alpha_j+\frac{3}{2}}^{\frac{\lambda_T^2}{\beta L_z^2}}
\right) \nonumber \\
&
+
\big(\delta_{d,2}+\delta_{d,1}\big)
\left(
\frac{2\pi\,\hslash^2\,\beta^{\frac{d}{2}-1}\,V^{d\textnormal{D}}}{\Gamma\left(\frac{d}{2}\right)\lambda^2\lambda_T^{d-2}\,\bar\theta L_x L_y}
\sum_jC\,\kappa_j\left(
\xi\,\mu\,I_{\alpha_j+\frac{d}{2}-1}^{\frac{\lambda_T^2}{\beta L_z^2}}+\widetilde\xi\,I_{\alpha_j+\frac{d}{2}}^{\frac{\lambda_T^2}{\beta L_z^2}}
\right)
\right. \nonumber \\
& \left. +
\frac{2\sqrt{\pi}\,\hslash^2\,\beta^{\frac{d-1}{2}}\,V^{d\textnormal{D}}}{\Gamma\left(\frac{d}{2}\right)\lambda^2\lambda_T^{d-1}\,\bar\theta L_x}
\sum_jG\,\kappa_j\left(
\xi\,\mu\,I_{\alpha_j+\frac{d-1}{2}}^{\frac{\lambda_T^2}{\beta L_y^2},\frac{\lambda_T^2}{\beta L_z^2}}+\widetilde\xi\,I_{\alpha_j+\frac{d+1}{2}}^{\frac{\lambda_T^2}{\beta L_y^2},\frac{\lambda_T^2}{\beta L_z^2}}
\right)
\right) \nonumber \\
& +
\delta_{d,1}
\frac{2\,\hslash^4\,\beta^{\frac{d}{2}}\,V^{d\textnormal{D}}}{\Gamma\left(\frac{d}{2}\right)\lambda^2\lambda_T^d\,\bar\theta}
\sum_j\kappa_j\left(
\xi\,\mu\,I_{\alpha_j+\frac{d}{2}}^{\frac{\lambda_T^2}{\beta L_x^2},\frac{\lambda_T^2}{\beta L_y^2}}+\widetilde\xi\,I_{\alpha_j+\frac{d}{2}+1}^{\frac{\lambda_T^2}{\beta L_x^2},\frac{\lambda_T^2}{\beta L_y^2}}
\right)
\ .
\end{align}

In the absence of a BEC, only the first sum in equations \eqref{Nirr} and \eqref{Hirr} contributes for large size, with $I_a^{\frac{\lambda_T^2}{\beta L_z^2}}$ replaced by $I_a^0$.
In the classical limit ($z=e^{\beta\mu}\ll1$ for ideal gases), one obtains $I_a^0=\mathcal{O}(z)$ from equation \eqref{polylog}, so that $\textnormal{Tr}(N\varrho_{\textnormal{irr}})$ and $\textnormal{Tr}(H\varrho_{\textnormal{irr}})$ scale as
$\mathcal{O}(Vz\ln z)$,
recalling the size scaling of the relaxation time \eqref{theta}.
These scalings in equations \eqref{WC.irr} and \eqref{QQ.irr} imply 
$W^{\textnormal{C}}_{k,\textnormal{irr}}=\mathcal{O}(Vz\ln^2 z)$
and
$W^{\textnormal{C}}_{k,\textnormal{irr}}-Q_{k,\textnormal{irr}}=\mathcal{O}(Vz\ln z)$.
In the quantum regime, the chemical potential is finite, the integrals $I_a^0$ are finite,
$\textnormal{Tr}(N\varrho_{\textnormal{irr}})$ and $\textnormal{Tr}(H\varrho_{\textnormal{irr}})$ scale as
$\mathcal{O}(V)$,
and consequently
$W^{\textnormal{C}}_{k,\textnormal{irr}}=\mathcal{O}(V)$ and $W^{\textnormal{C}}_{k,\textnormal{irr}}-Q_{k,\textnormal{irr}}=\mathcal{O}(V)$.
If the system is in a BEC phase (see appendix \ref{app.ideal}), then $\mu\simeq0$ and is much smaller than some of $\lambda_T^2/L^2_{x,y,z}$, and the integrals $I_a^A$ must be evaluated for both $A=\lambda_T^2/(\beta L_{x,y,z}^2)$ and $\mu$ approaching zero at large size:

\begin{equation}
I_a^A\underset{\substack{\mu\simeq0\\A\simeq0}}{\simeq}
\begin{cases}
\displaystyle \frac{1}{\beta^{a+1}}\,\Gamma(a+1)\,\zeta(a)+\frac{\pi a}{\sin(\pi a)}\,\frac{(-\mu)^{a-1}}{\beta^2}-\frac{1}{a+1}\,\frac{A^{a+1}}{\beta^2(A-\mu)^2} & \textnormal{if } 1<a<2 \\
\displaystyle -\frac{\ln(-\beta\mu)}{\beta^2}-\frac{1}{2\beta^2\left(1-\dfrac{\mu}{A}\right)^2} & \textnormal{if } a=1 \\
\displaystyle \frac{\pi a}{\sin(\pi a)}\,\frac{(-\mu)^{a-1}}{\beta^2}
-\frac{1}{a+1}\,\frac{A^{a+1}}{\beta^2(A-\mu)^2} & \textnormal{if } a<1 \textnormal{ and } -a\notin\mathbbm{N}^+ \\
\displaystyle \frac{1}{\beta^2\mu^2}\,\ln\left(\frac{\mu}{A}\right) & \textnormal{if } a=-1 \textnormal{ and } -\mu\gg A \\
\displaystyle \frac{1}{2\beta^2 A^2} & \textnormal{if } a=-1 \textnormal{ and } -\mu\ll A
\end{cases}
\ .
\end{equation}

Remind now that $\gamma_0\neq0$ (required for the uniqueness of the steady state as discussed at the end of section \ref{app.master}) and assume that $\gamma_0$ is finite. Under these conditions, one finds the series $\gamma_p^{-1}=\kappa+\sum_j\kappa_j\varepsilon_p^{\alpha_j}$ with $\kappa>0$ and $\alpha_j>0$ for all $j$: examples are the constant function $\gamma_p=\kappa$, the Lorentzian shapes $\gamma_p=1/(\kappa+\kappa'\varepsilon_p)$ or $h(\varepsilon_p)=1/\big(\sqrt{2\pi\kappa}+\kappa''\varepsilon_p\big)$, and the exponential functions $\gamma_p=e^{-\kappa'\varepsilon_p}/\kappa$ or $h(\varepsilon_p)=e^{-\kappa'\varepsilon_p}/\sqrt{2\pi\kappa}$.
In order to identify the leading terms in equations \eqref{Nirr} and \eqref{Hirr}, exploit the scaling of the box sizes and the chemical potential in BEC phases: $L_y\gtrsim e^{\alpha L_z}\textnormal{poly}(L_z)$ and $-\beta\mu\simeq e^{-f\rho L_z\lambda_T^2}\geqslant\mathcal{O}\big(\frac{\lambda_T^2}{L_y^2}\big)$ for 2D-BEC, $L_x\gtrsim\alpha' L_y L_z$ and $-\beta\mu\simeq\pi(f\rho L_y L_z\lambda_T)^{-2}\geqslant\mathcal{O}\big(\frac{\lambda_T^2}{L_x^2}\big)$ for 1D-BEC, $L_x\sim L_y\sim L_z$ and $-\beta\mu=(f\rho V)^{-1}$ for 0D-BEC (see appendix \ref{app.ideal}). Equation \eqref{Nirr} becomes

\begin{align}
\label{Nirr.BEC}
\textnormal{Tr}(N\varrho_{\textnormal{irr}})\simeq &
\begin{cases}
\displaystyle
-\frac{2\,\hslash^2\,\kappa}{\lambda^2\beta^2\mu\,\bar\theta}
\left(
\pi\,C\,\widetilde\xi
+2\sqrt{\pi}\,G
\left(
\xi+\frac{\lambda_T^2\,\widetilde\xi}{3\beta\mu L_y^2}
\right)
\right)
& \displaystyle \textnormal{2D-BEC} \\
\displaystyle
\frac{\hslash^2\,C\,\kappa\,\lambda_T}{\lambda^2\,\bar\theta L_y}
\sqrt{\frac{\pi^3}{-\beta^5\mu^3}}
\,
\big(\widetilde\xi-3\,\xi\big)
+\frac{8\,\hslash^2\,\kappa\,\xi\,\mu\,L_xL_y^3}{\sqrt{\pi}\,\lambda^2\lambda_T^4}
& \displaystyle \textnormal{1D-BEC} \\
\displaystyle \frac{2\,\hslash^2\,\xi}{\lambda^2\,\gamma_0\,\beta^2\mu\,\bar\theta}
& \displaystyle \textnormal{0D-BEC} \\
\end{cases} \nonumber \\
\simeq &
\begin{cases}
\displaystyle
\frac{2\,\hslash^2\kappa}{\lambda^2\beta\,\bar\theta}
\ e^{f\rho L_z\lambda_T^2}
\left(
\pi\,C\,\widetilde\xi
+2\sqrt{\pi}\,G
\left(
\xi-\frac{\widetilde\xi\,\lambda_T^2\,e^{f\rho\,L_z\lambda_T^2}}{3L_y^2}
\right)
\right)
& \displaystyle \textnormal{2D-BEC} \\
\displaystyle
\frac{\sqrt{\pi}\,\hslash^2\,\kappa}{\lambda^2\beta\,\bar\theta}
\left(
C\,\big(\widetilde\xi-3\,\xi\big)f^3\rho^3\lambda_T^4L_y^2L_z^3
-\frac{8\,\kappa\,\xi}{f^2\rho^2\lambda_T^6}
\ \frac{L_xL_y}{L_z^2}
\right)
& \displaystyle \textnormal{1D-BEC} \\
\displaystyle -\frac{2\,\hslash^2\,\xi\,f\rho\,V}{\lambda^2\,\gamma_0\,\beta\,\bar\theta}
& \displaystyle \textnormal{0D-BEC}
\end{cases}\ ,
\end{align}
where the first term of the 1D-BEC case dominates if $\alpha' L_yL_z\lesssim L_x<\mathcal{O}\big(L_yL_z^5\big)$ and the second term dominates when $L_x>\mathcal{O}\big(L_yL_z^5\big)$. The leading orders of equation \eqref{Hirr} are

\begin{align}
\label{Hirr.BEC}
\textnormal{Tr}(H\varrho_{\textnormal{irr}})\simeq &
\begin{cases}
\displaystyle
\frac{2\,\hslash^2\sqrt{\pi}\,\kappa}{\lambda^2\beta^2\,\bar\theta}
\left(
\frac{2\,G\,\widetilde\xi\,L_y}{5\,L_z}
+\left(\textnormal{terms} \propto\widetilde\xi\ln(-\beta\mu) \textnormal{ and} \propto\widetilde\xi\,V^0\right)
-\sqrt{\pi}\,C\,\xi
-\frac{G\,\xi\,\lambda_T^2}{3\beta\mu\,L_y^2}
\right)
& \displaystyle \textnormal{2D-BEC} \\
\displaystyle
\frac{3\pi\,\zeta\left(\frac{3}{2}\right)\hslash^2\,C\,\kappa\,\widetilde\xi\,L_z}{2\,\lambda^2\beta^2\lambda_T\,\bar\theta}
+\frac{\hslash^2\,\lambda_T\,C\,\kappa}{\lambda^2\,\bar\theta L_y}
\sqrt{\frac{\pi^3}{-\beta^5\mu}}
\,\big(\widetilde\xi-\xi\big)
+\frac{8\,\hslash^2\,\kappa\,\widetilde\xi L_x}{5\sqrt{\pi}\,\lambda^2\beta^2\,\bar\theta L_y}
+\frac{8\,\hslash^2\,\kappa\,\xi\mu L_x L_y}{3\sqrt{\pi}\,\lambda^2\beta \lambda_T^2\,\bar\theta}
& \displaystyle \textnormal{1D-BEC}
\\
\displaystyle
\frac{3\pi\,\zeta\left(\frac{3}{2}\right)\hslash^2\,C\,\kappa\,\widetilde\xi\,L_z}{2\,\lambda^2\beta^2\lambda_T\,\bar\theta}
+\frac{\hslash^2\,C\,\kappa\,L_z}{\lambda^2\lambda_T\,\bar\theta}
\sqrt{\frac{-\pi^3\mu}{\beta^3}}\big(3\,\widetilde\xi-\xi\big)
& \displaystyle \textnormal{0D-BEC} \\
\end{cases} \nonumber \\
\simeq &
\begin{cases}
\displaystyle
\frac{2\,\hslash^2\sqrt{\pi}\,\kappa}{\lambda^2\beta^2\,\bar\theta}
\left(
\frac{2\,G\,\widetilde\xi\,L_y}{5\,L_z}
+\left(\textnormal{terms} \propto\widetilde\xi\,L_z \textnormal{ and} \propto\widetilde\xi\,V^0\right)
-\sqrt{\pi}\,C\,\xi
+G\,\xi\,e^{f\rho\,L_z\lambda_T^2}\,\frac{\lambda_T^2}{3\,L_y^2}
\right)
& \displaystyle \textnormal{2D-BEC} \\
\displaystyle
\frac{\hslash^2\kappa}{\lambda^2\beta^2\,\bar\theta}
\left(
\frac{3\pi\,\zeta\left(\frac{3}{2}\right)C\,\widetilde\xi\,L_z}{2\,\lambda_T}
+\sqrt{\pi}\,C
\,\big(\widetilde\xi-\xi\big)f\rho\,\lambda_T^2 L_z
+\frac{8\,\widetilde\xi L_x}{5\sqrt{\pi}L_y}
-\frac{8\,\xi L_x}{3\sqrt{\pi}f^2\rho^2\lambda_T^4 L_y L_z^2}
\right) & \displaystyle \textnormal{1D-BEC} \\
\displaystyle
\frac{\pi\,\hslash^2\,\kappa}{\lambda^2\beta^2\,\bar\theta}
\left(
\frac{3\,\zeta\left(\frac{3}{2}\right)\,C\,\widetilde\xi\,L_z}{2\,\lambda_T}
+C\,\big(3\,\widetilde\xi-\xi\big)\sqrt{\frac{\pi^3L_z}{f\rho\,\lambda_T^2\,L_x L_y}}
\right)
& \displaystyle \textnormal{0D-BEC}
\end{cases}\ .
\end{align}
where the third term in the 1D-BEC phase dominates if $L_x>\mathcal{O}(L_yL_z)$ and the fourth term dominates the first two terms if $L_x>\mathcal{O}(L_yL_z^3)$. Higher order terms relevant in isothermal-isochoric transformations, where $\widetilde\xi=0$, are explicitely written.
\end{widetext}

The above expressions of $\textnormal{Tr}(N\varrho_{\textnormal{irr}})$ and $\textnormal{Tr}(H\varrho_{\textnormal{irr}})$, together with equations \eqref{Nirr.der.t} and \eqref{Hirr.der.t},
provide the size scaling of work and heat corrections in equation \eqref{WC.irr}, \eqref{WM.irr}, and \eqref{QQ.irr}. Using these scalings, one estimates the maximum power of irreversible cycles in the perturbative regime discussed here, the optimal times of each stroke and the efficiency at maximum power, as reported in appendix \ref{app.cycles} and in section \ref{irr}.

\section{Irreversible thermochemical cycles} \label{app.cycles}

In this section, the size scaling of the first order work and heat corrections in equations \eqref{WM.irr}, \eqref{WC.irr}, and \eqref{QQ.irr} are used to derive the size scalings of the optimal times $\tau_j^*$, of the maximum power $\pi_*$ and of the corresponding efficiency $\eta_*$ for irreversible thermochemical cycles.

\subsection{Irreversible cycles without BEC and in the classical limit} \label{app.irr.noBEC}

In the absence of BEC phases during the cycle, the computations in appendix \ref{en.corr} result in $W^{\textnormal{C}}_{k,\textnormal{irr}}=\mathcal{O}(V)$, and $W^{\textnormal{C}}_{k,\textnormal{irr}}-Q_{k,\textnormal{irr}}=\mathcal{O}(V)$ for all $k$.
The load of the reversible cycles was proven to be extensive $W^{\textnormal{M}}_{\textnormal{rev}}=\mathcal{O}(V)$ in section \ref{engines}.
These scalings imply that
the optimal times in equation \eqref{tau.max} for every stroke
have the same size scaling of the relaxation time $\bar\theta=\mathcal{O}\big(L_z/V\big)$.
The output power is therefore maximised, within the perturbative regime $\tau_j^*\gg\bar\theta$, for the scaling $\tau_j^*=\mathcal{O}\big(\frac{L_z}{sV}\big)$ with $s\ll1$.
Consequently, the maximum power scales as $\pi_*=\mathcal{O}\big(\frac{s\,V^2}{L_z}\big)$, the corresponding load as $W^{\textnormal{M}}_*=W^{\textnormal{M}}_{\textnormal{rev}}+\mathcal{O}(s\,V)$, and the efficiency as $\eta_*=\eta_{\textnormal{rev}}+\mathcal{O}(s\,V^0)$.

The size scalings in the classical limit are obtained from those without BECs imposing the condition \eqref{class}. This condition for ideal gases reads $\mathcal{N}\simeq z_{\textnormal{cl}}V/\lambda_T^3$ with small fugacity $z_{\textnormal{cl}}=e^{\beta\mu}\ll1$ (see appendices \ref{app.class} and \ref{app.ideal}).
The classical limit and the first law of thermodynamics for reversible isothermal cycles then imply
$W_{\textnormal{rev}}^{(M)}=-W_{\textnormal{rev}}^{(C)}=\mathcal{O}(z_{\textnormal{cl}}V)$,
while irreversible energy corrections are
$W^{\textnormal{C}}_{k,\textnormal{irr}}=\mathcal{O}(V z_{\textnormal{cl}}\ln^2 z_{\textnormal{cl}})$
and
$W^{\textnormal{C}}_{k,\textnormal{irr}}-Q_{k,\textnormal{irr}}=\mathcal{O}(V z_{\textnormal{cl}}\ln z_{\textnormal{cl}})$.
The optimal time in equation \eqref{tau.max} is $\tau_{\textnormal{class}}=\mathcal{O}\big(\frac{L_z}{V}|\ln z_{\textnormal{cl}}|\big)$, the maximum power is $\pi_{\textnormal{class}}=\mathcal{O}\big(\frac{z_{\textnormal{cl}}\,V^2}{|\ln z_{\textnormal{cl}}|L_z}\big)$, and the efficiency at maximum power
scales as $\eta_{\textnormal{class}}=\mathcal{O}(z_{\textnormal{cl}})$.

In the cycles discussed in section \ref{engines}, $\eta_{\textnormal{rev}}$ is finite and approaches one close to BEC transitions, and thus the quantum regime even without BECs exhibits much higher efficiency than the classical limit and larger power if $s>z_{\textnormal{cl}}/|\ln z_{\textnormal{cl}}|$.

\subsection{Irreversible cycles with BECs}

Different BEC phases of the working substance emerge at different confinement anisotropies.
The three cases considered in section \ref{irr.cycles.BEC} for $d$D-BECs are:

\begin{itemize}
\item[$d=2$:] $L_z=\mathcal{O}(\ln V)$ and $L_x\simeq L_y=\mathcal{O}\big(\sqrt{V/\ln V}\big)$,
\item[$d=1$:] $L_x=\mathcal{O}\big(V^{\frac{\chi}{\chi+1}}\big)$ and $L_y\sim L_z=\mathcal{O}\big(V^{\frac{1}{2\chi+2}}\big)$ with $\chi\geqslant 1$, i.e., $L_x=\mathcal{O}(L_y L_z)^\chi$,
\item[$d=0$:] $L_x\sim L_y\sim L_z=\mathcal{O}\big(\sqrt[3]{V}\big)$.
\end{itemize}

The size scaling of the work and heat corrections in equations
\eqref{WC.irr}, \eqref{WM.irr}, and \eqref{QQ.irr}
depend on the function $h(\varepsilon)$, and are explicitly computed in appendix \ref{en.corr} assuming $h(\varepsilon_p)\neq0$ (or $\gamma_p\neq0$) and finite $h(0)$. The condition $h(\varepsilon_p)\neq0$ guarantees that the grandcanonical ensemble is the unique steady state at every time, otherwise also the instantaneous Hamiltonian eigenstates with energy $\varepsilon_p$ such that $h(\varepsilon_p)=0$ are steady states. This large class of functions contains the constant function, the exponential decay, and the Lorentzian.

The result for the corrections to the chemical work in equation
\eqref{WC.irr}
is

\begin{equation}
\label{corrWc0}
W^{\textnormal{C}}_{k,\textnormal{irr}}=
\begin{cases}
\displaystyle
\mathcal{O}\left(\frac{V}{L_z}\right)
& \displaystyle \textnormal{2D-BEC} \\
\displaystyle
\mathcal{O}(V)+\mathcal{O}\left(\frac{L_x^2}{L_z^4}\right)
& \displaystyle \textnormal{1D-BEC} \\
\displaystyle
\mathcal{O}\left(\frac{V}{L_z}\right)
& \displaystyle \textnormal{0D-BEC}
\end{cases}\ ,
\end{equation}
where the first (second) term for the 1D-BEC dominates if $L_x<\mathcal{O}\big(L_y L_z^5\big)$ ($L_x>\mathcal{O}\big(L_y L_z^5\big)$).
The other energy corrections used in the computation of \eqref{tau.max}, namely $W^{\textnormal{C}}_{k,\textnormal{irr}}-Q_{k,\textnormal{irr}}$, have different size scalings for isothermal-isochoric transformations compared to the other strokes. Indeed, the leading orders for all other strokes are

\begin{equation}
\label{corrWcmQ}
W^{\textnormal{C}}_{k,\textnormal{irr}}-Q_{k,\textnormal{irr}}=
\begin{cases}
\displaystyle
\mathcal{O}\left(\frac{L_xL_y^2}{L_z}\right)
& \displaystyle \textnormal{2D-BEC} \\
\displaystyle
\mathcal{O}\big(L_x^2\big)
& \displaystyle \textnormal{1D-BEC} \\
\displaystyle
\mathcal{O}(V)
& \displaystyle \textnormal{0D-BEC}
\end{cases}\ .
\end{equation}
Neverthless, these size scalings originate from orders that multiply $\widetilde\xi$, namely the time derivative of the temperature or that of the volume, and these contributions vanish for isothermal-isochoric processes.
The size scaling for isothermal-isochoric transformations, resulting by computations in appendix \ref{en.corr} with $\widetilde\xi=0$, is

\begin{equation}
\label{corrWcmQisoTV}
W^{\textnormal{C}}_{k,\textnormal{irr}}-Q_{k,\textnormal{irr}}=
\begin{cases}
\displaystyle
\mathcal{O}\left(\frac{V}{L_z}\right)
& \displaystyle \textnormal{2D-BEC} \\
\displaystyle
\mathcal{O}\left(V\right)
+\mathcal{O}\left(\frac{L_x^2}{L_z^2}\right)
& \displaystyle \textnormal{1D-BEC} \\
\displaystyle
\mathcal{O}\big(\sqrt{V}\big)
& \displaystyle \textnormal{0D-BEC} \\
\end{cases}\ ,
\end{equation}
with the first (second) term of the 1D-BEC dominating when $L_x<\mathcal{O}\big(L_y L_z^3\big)$ ($L_x>\mathcal{O}\big(L_y L_z^3\big)$).

The isothermal chemical Carnot and Otto cycles are treated separetely in the following, because isothermal-isochoric transformations are part only of the latter.

\subsubsection{Irreversible chemical Carnot cycle with a BEC during the third stroke}

If the system is a BEC only when chemical work is released (during the third stroke),
the reversible load \eqref{load.carnot} is extensive $W^{\textnormal{M}}_{\textnormal{rev}}=\mathcal{O}(V)$, such that the scaling \eqref{corrWcmQ} and that without BECs imply

\begin{equation}
\label{tau.scaling.BEC.not3}
\tau_{1,2,4}^*=
\begin{cases}
\displaystyle
\mathcal{O}\left(\frac{1}{\sqrt[4]{V^3\ln V}}\right)
& \displaystyle \textnormal{2D-BEC} \\
\displaystyle
\mathcal{O}\left(\frac{1}{s\,V^{\frac{3}{4}}}\right)
& \displaystyle \textnormal{1D-BEC, } \chi=1 \\
\displaystyle
\mathcal{O}\left(\frac{1}{V^{\frac{\chi+2}{2\chi+2}}}\right)
& \displaystyle \textnormal{1D-BEC, } \chi>1 \\
\displaystyle
\mathcal{O}\left(\frac{1}{s\,V^\frac{2}{3}}\right)
& \displaystyle \textnormal{0D-BEC}
\end{cases}\ ,
\end{equation}
and

\begin{equation}
\label{tau.scaling.BEC.3}
\tau_3^*=
\begin{cases}
\displaystyle
\mathcal{O}\left(\frac{1}{\sqrt{V(\ln V)^3}}\right)
& \displaystyle \textnormal{2D-BEC} \\
\displaystyle
\mathcal{O}\left(\frac{1}{s\,V^{\frac{3}{4}}}\right)
& \displaystyle \textnormal{1D-BEC, } \chi=1 \\
\displaystyle
\mathcal{O}\left(\frac{1}{V^{\frac{3}{2\chi+2}}}\right)
& \displaystyle \textnormal{1D-BEC, } \chi>1 \\
\displaystyle
\mathcal{O}\left(\frac{1}{s\,V^\frac{2}{3}}\right)
& \displaystyle \textnormal{0D-BEC}
\end{cases}\ .
\end{equation}
The factor $s\ll1$ has been introduced in order to remain in the perturbative regime, when the optimal times in equation \eqref{tau.max} scale as the relaxation time $\bar\theta=\mathcal{O}(L_z/V)$,
as discussed in section \ref{app.irr.noBEC}.

Using the scalings of $W^{\textnormal{M}}_{\textnormal{rev}}$, $\tau_j^*$ and \eqref{corrWc0}, one derives the efficiency at maximum power and the maximum power normalised to the classical limit of section \ref{irr.Carnot3}.

\subsubsection{Irreversible chemical Carnot cycle with a BEC during all the strokes} \label{irr.Carnot.all}

If the system is in the same BEC phase during the entire chemical Carnot cycle, the load in the reversible limit is sub-extensive: applying the above scalings of the chemical potential and of the box size $L_{x,y,z}$ to the work done during the entire chemical Carnot cycle (see section \ref{Carnot}) at finite density, one obtains

\begin{equation} \label{WMrev.sub}
W_{\textnormal{rev}}^{\textnormal{M}}=(\mu_3-\mu_1)\left(\mathcal{N}_1-\mathcal{N}_3\right)=
\begin{cases}
\displaystyle
\mathcal{O}\big(\ln V\big)
& \displaystyle \textnormal{2D-BEC} \\
\displaystyle
\mathcal{O}\left(V^{\frac{\chi-1}{\chi+1}}\right)
& \displaystyle \textnormal{1D-BEC} \\
\displaystyle
\mathcal{O}\left(V^0\right)
& \displaystyle \textnormal{0D-BEC}
\end{cases}\ ,
\end{equation}
where it has been assumed in the 2D-BEC case that the chemical potential saturates its lower bound $-\beta\mu\geqslant\mathcal{O}\big(\frac{\lambda_T^2}{L_y^2}\big)$ which provides a lower bound for the scaling of $W_{\textnormal{rev}}^{\textnormal{M}}$ and consequently an upper bound for the times $\tau_j^*$ (see equations \eqref{tau.max}).
Therefore, the time needed for every stroke is

\begin{equation}
\label{tau.scaling.BEC2}
\tau_j^*=
\begin{cases}
\displaystyle
\mathcal{O}\left(\sqrt{\frac{V}{\big(\ln V\big)^5}}\right)
& \displaystyle \textnormal{2D-BEC} \\
\displaystyle
\mathcal{O}\left(V^{\frac{1}{2\chi+2}}\right)
& \displaystyle \textnormal{1D-BEC} \\
\displaystyle
\mathcal{O}\big(V^\frac{1}{3}\big)
& \displaystyle \textnormal{0D-BEC}
\end{cases}\ .
\end{equation}
The efficiency at maximum power is

\begin{equation}
\label{eta.max.BEC2}
\eta_*=\frac{\eta_{\textnormal{rev}}}{2}+
\begin{cases}
\displaystyle
\mathcal{O}\left(\sqrt{\frac{\big(\ln V\big)^3}{V}}\right)
& \displaystyle \textnormal{2D-BEC} \\
\displaystyle
\mathcal{O}\left(\frac{1}{V^{\frac{\chi-1}{\chi+1}}}\right)
& \displaystyle \textnormal{1D-BEC, } 1\leqslant\chi\leqslant 3 \\
\displaystyle
\mathcal{O}\left(\frac{1}{V^{\frac{2}{\chi+1}}}\right)
& \displaystyle \textnormal{1D-BEC, } \chi\geqslant 3 \\
\displaystyle
\mathcal{O}\left(\frac{1}{V^\frac{1}{3}}\right)
& \displaystyle \textnormal{0D-BEC}
\end{cases}\ .
\end{equation}

As discussed in section \ref{Carnot}, the efficiency $\eta_{\textnormal{rev}}$ is finite when the substance is always in the same BEC phase and approaches $1$ for 2D-BEC and large size. Therefore the efficiency at maximum power in \eqref{eta.max.BEC2} is again much larger than the classical limit $\eta_{\textnormal{class}}=\mathcal{O}(z_{\textnormal{cl}})\ll1$. Nevertheless, the subextensive load \eqref{WMrev.sub} and the scaling of optimal times \eqref{tau.scaling.BEC2}
imply small output power.

\subsubsection{Irreversible chemical Otto cycle with a BEC during the third stroke}

In the chemical Otto cycle, the first and the third strokes are isothermal-isochoric processes, and one has to consider the corrections \eqref{corrWcmQisoTV} instead of \eqref{corrWcmQ}.
If the system is a BEC only during the third stroke, as discussed in section \ref{Otto},
the load is extensive, $W^{\textnormal{M}}_{\textnormal{rev}}=\mathcal{O}(V)$. Therefore, the optimal times for each stroke are

\begin{equation}
\label{tau.scaling.BEC.not3Otto}
\tau_{1,2,4}^*=
\begin{cases}
\displaystyle
\mathcal{O}\left(\frac{\ln V}{s\,V}\right)
& \displaystyle \textnormal{2D-BEC} \\
\displaystyle
\mathcal{O}\left(\frac{1}{s\,V^{\frac{2\chi+1}{2\chi+2}}}\right)
& \displaystyle \textnormal{1D-BEC, } \chi\leqslant 2 \\
\displaystyle
\mathcal{O}\left(\frac{1}{V^{\frac{\chi+3}{2\chi+2}}}\right)
& \displaystyle \textnormal{1D-BEC, } \chi\geqslant 2 \\
\displaystyle
\mathcal{O}\left(\frac{1}{s\,V^\frac{2}{3}}\right)
& \displaystyle \textnormal{0D-BEC}
\end{cases}\ ,
\end{equation}
and

\begin{equation}
\label{tau.scaling.BEC.3Otto}
\tau_3^*=
\begin{cases}
\displaystyle
\mathcal{O}\left(\frac{\ln V}{s\,V}\right)
& \displaystyle \textnormal{2D-BEC} \\
\displaystyle
\mathcal{O}\left(\frac{1}{s\,V^{\frac{2\chi+1}{2\chi+2}}}\right)
& \displaystyle \textnormal{1D-BEC, } \chi\leqslant 2 \\
\displaystyle
\mathcal{O}\left(\frac{1}{V^{\frac{5}{2\chi+2}}}\right)
& \displaystyle \textnormal{1D-BEC, } \chi\geqslant 2 \\
\displaystyle
\mathcal{O}\left(\frac{1}{V^\frac{11}{12}}\right)
& \displaystyle \textnormal{0D-BEC}
\end{cases}\ ,
\end{equation}
Remind that, when the times in equation \eqref{tau.max} do not
fulfil the condition $\tau_j^*\gg\bar\theta$,
the optimal times within the perturbative regime are $\tau_j^*=\bar\theta/s=\mathcal{O}\big(\frac{L_z}{sV}\big)$ with $s\ll1$.

Using the scalings derived so far, one obtains the efficiency at maximum power and the maximum power normalised to the classical limit of section \ref{irr.Otto3}.

\subsubsection{Irreversible chemical Otto cycle with a BEC during all the strokes} \label{irr.Otto.all}

When the system is in the same BEC phase during the entire cycle, equations \eqref{WC1.0D-BEC}, \eqref{WC3.0D-BEC}, \eqref{WC1.1D-BEC}, \eqref{WC3.1D-BEC}, \eqref{WC1.2D-BEC}, \eqref{WC3.2D-BEC} imply that the reversible load,
$W_{\textnormal{rev}}^{\textnormal{M}}=-W_{1,\textnormal{rev}}^{\textnormal{C}}-W_{3,\textnormal{rev}}^{\textnormal{C}}$ scales as

\begin{equation} \label{WMrev.sub.Otto}
W_{\textnormal{rev}}^{\textnormal{M}}=
\begin{cases}
\displaystyle
\mathcal{O}\big(V^0\big)
& \displaystyle \textnormal{2D-BEC} \\
\displaystyle
\mathcal{O}\left(V^{\frac{\chi-1}{\chi+1}}\right)
& \displaystyle \textnormal{1D-BEC} \\
\displaystyle
\mathcal{O}\left(V^0\right)
& \displaystyle \textnormal{0D-BEC}
\end{cases}\ ,
\end{equation}
Therefore, the times needed for each stroke are

\begin{equation}
\label{tau.scaling.BEC2.13Otto}
\tau_{1,3}^*=
\begin{cases}
\displaystyle
\mathcal{O}\left(\frac{V}{\big(\ln V\big)^3}\right)^\frac{1}{4}
& \displaystyle \textnormal{2D-BEC} \\
\displaystyle
\mathcal{O}\left(V^{\frac{2-\chi}{2\chi+2}}\right)
& \displaystyle \textnormal{1D-BEC, } \chi\leqslant 2 \\
\displaystyle
\mathcal{O}\left(V^0\right)
& \displaystyle \textnormal{1D-BEC, } \chi\geqslant 2 \\
\displaystyle
\mathcal{O}\big(V^\frac{1}{12}\big)
& \displaystyle \textnormal{0D-BEC}
\end{cases}
\end{equation}
and

\begin{equation}
\label{tau.scaling.BEC2.24Otto}
\tau_{2,4}^*=
\begin{cases}
\displaystyle
\mathcal{O}\left(\sqrt{\frac{V}{\big(\ln V\big)^3}}\right)
& \displaystyle \textnormal{2D-BEC} \\
\displaystyle
\mathcal{O}\left(V^{\frac{1}{2\chi+2}}\right)
& \displaystyle \textnormal{1D-BEC} \\
\displaystyle
\mathcal{O}\big(V^\frac{1}{3}\big)
& \displaystyle \textnormal{0D-BEC}
\end{cases}\ .
\end{equation}

The efficiency at maximum power is

\begin{equation}
\label{eta.max.BEC2Otto}
\eta_*=\frac{\eta_{\textnormal{rev}}}{2}+
\begin{cases}
\displaystyle
\mathcal{O}\left(\frac{\big(\ln V\big)^3}{V}\right)^\frac{1}{4}
& \displaystyle \textnormal{2D-BEC} \\
\displaystyle
\mathcal{O}\left(\frac{1}{V^{\frac{\chi-1}{2\chi+2}}}\right)
& \displaystyle \textnormal{1D-BEC, } 1\leqslant\chi\leqslant 2 \\
\displaystyle
\mathcal{O}\left(\frac{1}{V^{\frac{2\chi-3}{2\chi+2}}}\right)
& \displaystyle \textnormal{1D-BEC, } 2\leqslant\chi\leqslant 3 \\
\displaystyle
\mathcal{O}\left(\frac{1}{V^{\frac{3}{2\chi+2}}}\right)
& \displaystyle \textnormal{1D-BEC, } \chi\geqslant 3 \\
\displaystyle
\mathcal{O}\left(\frac{1}{V^\frac{1}{12}}\right)
& \displaystyle \textnormal{0D-BEC}
\end{cases}\ .
\end{equation}
As for the chemical Carnot cycle, the efficiency at maximum power in \eqref{eta.max.BEC2Otto} is again much larger than the classical limit $\eta_{\textnormal{class}}=\mathcal{O}(z_{\textnormal{cl}})\ll1$, but with small output power because of the subextensive load \eqref{WMrev.sub.Otto} and of the scaling of optimal times \eqref{tau.scaling.BEC2.13Otto} and \eqref{tau.scaling.BEC2.24Otto}.

{\bf Acknowledgements} \\
U.~M. is financially supported by the European Union's Horizon 2020 research and innovation programme under the Marie Sk\l odowska-Curie grant agreement No. 754496 - FELLINI.




%


\end{document}